\documentclass[11pt,leqno]{amsart}
\usepackage{amsmath,epsfig}
\newcommand{\levy}{L\'evy}

\newcommand{\clt}{central limit theorem}

\newcommand{\garch}{{\rm GARCH}$(1,1)$}

\newcommand{\asy}{asymptotic}

\newcommand{\ts}{time series}

\newcommand{\garchpq}{{\rm GARCH}$(p,q)$}

\newtheorem{lemma}{Lemma}[section]

\newtheorem{theorem}[lemma]{Theorem}

\newtheorem{proposition}[lemma]{Proposition}
\newtheorem{definition}[lemma]{Definition}
\newtheorem{corollary}[lemma]{Corollary}
\newtheorem{example}[lemma]{Example}
\newtheorem{exercise}[lemma]{Exercise}
\newtheorem{remark}[lemma]{Remark}
\newtheorem{fig}[lemma]{Figure}
\newtheorem{tab}[lemma]{Table}

\newcommand{\cid}{\stackrel{d}{\rightarrow}}

\newcommand{\bth}{\begin{theorem}}
\newcommand{\ethe}{\end{theorem}}

\newcommand{\bre}{\begin{remark}\em }
\newcommand{\ere}{\end{remark}}

\newcommand{\ble}{\begin{lemma}}
\newcommand{\ele}{\end{lemma}}

\newcommand{\bde}{\begin{definition}}
\newcommand{\ede}{\end{definition}}
\newcommand{\bco}{\begin{corollary}}
\newcommand{\eco}{\end{corollary}}

\newcommand{\bpr}{\begin{proposition}}
\newcommand{\epr}{\end{proposition}}

\newcommand{\bexer}{\begin{exercise}}
\newcommand{\eexer}{\end{exercise}}

\newcommand{\bexam}{\begin{example}}
\newcommand{\eexam}{\end{example}}

\newcommand{\bfi}{\begin{fig}}
\newcommand{\efi}{\end{fig}}

\newcommand{\btab}{\begin{tab}}
\newcommand{\etab}{\end{tab}}

\newcommand{\fidi}{finite-dimensional distribution}
\newcommand{\rv}{random variable}

\newcommand{\var}{{\rm var}}

\newcommand{\cov}{{\rm cov}}

\newcommand{\rhs}{right-hand side}
\newcommand{\df}{distribution function}

\newcommand{\beao}{\begin{eqnarray*}}
\newcommand{\eeao}{\end{eqnarray*}\noindent}

\newcommand{\beam}{\begin{eqnarray}}
\newcommand{\eeam}{\end{eqnarray}\noindent}

\newcommand{\beqq}{\begin{equation}}
\newcommand{\eeqq}{\end{equation}\noindent}

\newcommand{\bce}{\begin{center}}
\newcommand{\ece}{\end{center}}

\newcommand{\barr}{\begin{array}}
\newcommand{\earr}{\end{array}}

\newcommand{\stp}{\stackrel{P}{\rightarrow}}
\newcommand{\std}{\stackrel{d}{\rightarrow}}

\newcommand{\stv}{\stackrel{v}{\rightarrow}}
\newcommand{\stw}{\stackrel{w}{\rightarrow}}

\newcommand{\vague}{\stackrel{\lower0.2ex\hbox{$\scriptscriptstyle
                    \it{v} $}}{\rightarrow}}
\newcommand{\weak}{\stackrel{\lower0.2ex\hbox{$\scriptscriptstyle
                    \it{w} $}}{\rightarrow}}
\newcommand{\what}{\stackrel{\lower0.2ex\hbox{$\scriptscriptstyle
                    \it{\hat{w}} $}}{\rightarrow}}

\newcommand{\bdis}{\begin{displaymath}}
\newcommand{\edis}{\end{displaymath}\noindent}

\newcommand{\nto}{n\to\infty}

\newcommand{\xto}{x\to\infty}

\newcommand{\ov}{\overline}
\newcommand{\wt}{\widetilde}
\newcommand{\wh}{\widehat}

\newcommand{\regvary}{regularly varying}

\newcommand{\regvar}{regular variation}

\newcommand{\bbr}{{\mathbb R}}

\newcommand{\bbz}{{\mathbb Z}}

\newcommand{\bbs}{{\mathbb S}}

\newcommand{\con}{convergence}

\newcommand{\st}{such that}

\newcommand{\wrt}{with respect to}

\newcommand{\fct}{function}

\newcommand{\ds}{distribution}

\newcommand{\rep}{representation}

\newcommand{\seq}{sequence}

\newcommand{\pro}{probabilit}

\newcommand{\ms}{measure}

\newcommand{\bfz}{{\bf z}}
\newcommand{\bfS}{{\bf S}}

\begin{document}

\bibliographystyle{plain}
\title[The extremogram]{
Estimating Extremal Dependence in Univariate and Multivariate Time Series
Via the Extremogram}
\address{Department of Statistics,
Columbia University,
1255 Amsterdam Ave.
New York, NY 10027, U.S.A.}
\email{rdavis@stat.columbia.edu\,, www.stat.columbia.edu/$\sim$rdavis}
\address{Department of Mathematics,
University of Copenhagen,
Universitetsparken 5,
DK-2100 Copenhagen,
Denmark}
\email{mikosch@math.ku.dk\,, www.math.ku.dk/$\sim$mikosch}
\address{Department of Statistics,
Columbia University,
1255 Amsterdam Ave.
New York, NY 10027, U.S.A.}
\email{ivor@stat.columbia.edu}

\maketitle
\begin{center}Richard A. Davis\footnote{coordinating author: Department of Statistics, 1255 Amsterdam Avenue, Columbia University, New York, NY 10027, USA; email: rdavis@stat.columbia.edu
 {\it Primary JEL Classification Code:} C50. }, Columbia University \end{center}
\begin{center}Thomas Mikosch, University of Copenhagen \end{center}
\begin{center}Ivor Cribben, Columbia University \end{center}

\bigskip
\noindent{\bf Abstract.}
Davis and Mikosch \cite{davis:mikosch:2009} introduced the
extremogram as a flexible quantitative tool for measuring
various types of extremal dependence in a stationary time
series. There we showed some standard statistical properties of the
sample extremogram. A major difficulty was the construction of
credible confidence bands for the extremogram. In this paper, we
employ the stationary
bootstrap to overcome this problem. Moreover, we introduce the
cross extremogram as a \ms\ of extremal serial dependence between two
or more time series. We also study the extremogram for return times
between extremal events. The use of the stationary
bootstrap for the extremogram and the resulting
interpretations are illustrated in several
univariate and multivariate financial time series examples.

\newpage
\baselineskip=15pt
\section{Introduction}\setcounter{equation}{0}

With the wild swings recently seen in the financial markets and climatic conditions, there has been renewed interest in understanding and modeling extreme events. The extremogram, developed in Davis and Mikosch \cite{davis:mikosch:2009}, is a flexible tool that provides a quantitative measure of dependence of extreme events in a stationary time series. In many respects, one can view the extremogram as the extreme-value analog of the autocorrelation function (ACF) of a stationary process. In classical time series modeling the ACF, and its sample counterpart, are the workhorses for measuring and estimating linear dependence in the family of linear time series processes. While the ACF has some use in measuring dependence in non-linear time series models, especially when applied to non-linear functions of the data such as absolute values and squares, it has limited value in assessing dependence between extreme events. On the other hand, the extremogram only considers observations, or groups of observations, which are large.

For a $d$-dimensional strictly stationary time series $(X_t)$, the {\em extremogram} is defined for two sets $A$ and $B$ bounded away from 0 by\footnote{A set $C$ is bounded away from zero if $C\subset\{y:\,|y|>r\}$ for some $r>0$.}
\beam \label{eq:ext}
\rho_{A,B}(h)=\lim_{x\to\infty}P(x^{-1}X_h\in B\mid x^{-1}X_0\in A),\quad h=0,1,2,\ldots,
\eeam
provided the limit exists.
Since $A$ and $B$ are  bounded
away from zero, the events $\{x^{-1}X_0\in A\}$ and $\{x^{-1}X_h\in B\}$ are becoming extreme in the sense the probabilities of these events are converging to zero with $x\to\infty$. In the special
case of a univariate time series { and the choice of the sets}
$A=B=(1,\infty)$, the extremogram reduces to the (upper) tail
dependence coefficient between $X_0$ and $X_h$ that is often used in
extreme value theory { and quantitative risk managament; see e.g. McNeil
et al. \cite{mcneil:frey:embrechts:2005}.}  In this case, one is
interested in computing the impact of a large value of the time series
on a future value $h$ time-lags ahead.  With creative choices of $A$
and $B$, one can investigate interesting sources of extremal
dependence that may arise not only in the upper and lower tails, but
also in other extreme regions of the sample space; see Sections
\ref{sec:bootext} and \ref{sec:further} for some examples.
\par
{ We
  would like to emphasize that the extremogram is a {\em conditional} \ms\
  of extremal serial dependence. Therefore it is particularly suited for
  financial applications where one is often interested in the
  persistence of a shock (an extremal event on the stock market say)
at future instants of time. Another good reason for using the extremogram
for financial \ts\ is a statistical one: for large $x$, the quantities
$P(x^{-1}X_h\in B\mid x^{-1}X_0\in A)$ are rare event \pro ies; their
non-parametric estimation cannot be based on standard empirical
process techniques and requires large sample sizes. Fortunately, long financial
time series are available and therefore the study of their extremal
serial behavior is not only desirable but also  possible. Financial \ts\ often have the
(from a statistical point of view) desirable property that they are
heavy-tailed, i.e. extreme large and small values
are rather pronounced and occur in clusters. The extremogram and its
modifications discussed in this paper allow one to give clear
quantititave descriptions of the size  and persistence of such clusters.}

In estimating the extremogram, the limit on $x$ in \eqref{eq:ext} is
replaced by a high quantile $a_m$ of the process.
Defining $a_m$ as the $(1-1/m)$-quantile of the stationary distribution of $|X_t|$,
the sample extremogram based on the observations $X_1,\ldots,X_n$ is given by
\beam\label{eq:sext}
\wh\rho_{A,B}(h)=\frac{\sum_{t=1}^{n-h}I_{\{a_m^{-1}X_{t+h}\in B, a_m^{-1}X_t\in
    A\}}}{\sum_{t=1}^{n}I_{\{a_m^{-1}X_t\in A\}}}\,.
\eeam
In order to have a consistent result, we require
$m=m_n\to \infty$ with $m/n\to 0$ as $n\to \infty$.
In practice, we do not know $a_m$
and therefore it has to be replaced by a corresponding empirical
quantile, i.e., by one of the largest observations.
Under suitable mixing conditions and
other distributional assumptions that ensure the limit in \eqref{eq:ext} exists, it was shown in Davis and Mikosch \cite{davis:mikosch:2009} that
$\wh\rho_{A,B}(h)$ is asymptotically normal; i.e.,
\beam \label{eq:extnormal}
\sqrt{n/m}\,(\wh\rho_{A,B}(h)-\rho_{A,B:m}(h))\cid N(0,\sigma^2_{A,B}(h)),
\eeam
where
\beam \label{eq:pext}
\rho_{A,B:m}(h)=P(a_m^{-1}X_h \in B\mid a_m^{-1}X_0\in A)\,.
\eeam
We refer to \eqref{eq:pext}
as the pre-asymptotic extremogram (PA-extremogram).
\par
There are several obstacles
in directly applying \eqref{eq:extnormal} for constructing confidence bands
for the extremogram:
\begin{enumerate}
\item[(i)] The asymptotic variance
$\sigma^2_{A,B}(h)$ is based on an infinite sum of unknown quantities
and typically does not have a closed-form expression.
\item[(ii)] Estimating
$\sigma^2_{A,B}(h)$ is similar to estimating the
{ \asy\ }variance of a sample mean from a
time series and is often difficult in practice.
\item[(iii)]
The PA-extremogram cannot always be replaced by its limit.
\end{enumerate}
For (i) and (ii), we turn to bootstrap procedures to approximate the distribution of $(\hat\rho_{A,B}(h) -\rho_{A,B:m}(h))$.  This will allow us to construct {\it credible} (asymptotically correct) confidence bands for the PA-extremogram.  As for (iii), a non-parametric bootstrap does not allow us to overcome the bias concern.  We note, however, that the PA-extremogram is a conditional probability that is often the quantity of primary interest in applications.  That is, one is typically interested in estimating conditional probabilities of {\it extreme} events as a measure of extremal dependence so that it is not necessary, and perhaps not even desirable to replace the PA-extremogram with the extremogram in \eqref{eq:extnormal}.

The objective of this paper is to apply the bootstrap to the sample extremogram in order to overcome these limitations. By now, there are many non-parametric bootstrap procedures in the literature that are designed for use with stationary time series.  Many of these involve some form of resampling from blocks of observations.  That is, in constructing a bootstrap replicate of the time series,  long stretches of the time series are stitched together in order to replicate the joint distributions of the process.  While for a finite sample size $n$, it is impossible to replicate all the joint distributions, we can only sample from  at most the $m$-variate distributions (for $m<n$) by sampling blocks of $m$ consecutive observations.  In order to obtain consistency of the procedure, $m$ is allowed to grow  with $n$ at a suitable rate.
In this paper, we adopt the stationary bootstrap approach as
described in Politis and Romano \cite{politis:romano:1994} in which the block sizes are given by
independent geometric random variables.
Since the blocks are of random length,
the stationary bootstrap is useful as an exploratory
device in which dependence beyond a fixed block
length can be discovered.
\par
In our case, there are significant differences in the extremogram setting of our bootstrap application from the traditional one.  First, the summands in the numerator and
denominator of \eqref{eq:sext}  form a triangular array of random
variables and cannot be cast as a single stationary sequence.  Second,
most bootstrapping applications in extreme value theory, even in the
iid case, require the replicate time series to be of smaller order
than the original sample size $n$; see e.g. Section 6.4 of Resnick \cite{resnick:2007}.  On the other hand, we are able to overcome these drawbacks and show that the bootstrapped sample extremogram, based on the replicates of size $n$ provides an asymptotically correct approximation to the left-hand side of \eqref{eq:extnormal} provided the blocks grow at a proper rate.
\par
In addition to providing a non-parametric estimate of the nature of
extremal dependence as a function of time-lag, the extremogram can
also provide valuable guidance in various phases of the typical time
series modeling paradigm. For example, the sample extremogram might
provide insight into the choice of models for the data with the goal
of delivering models that are compatible with the extremal dependence.
In the standard approach, models are often selected to fit the center
of the distribution and can be inadequate for describing the extremes
in the data. On the other hand, if the primary interest is on modeling
extremes, then the modeling exercise should focus on this aspect. The quality
of fit could be judged by assessing compatibility of the sample extremogram
with the fitted model extremogram. Moreover, the sample extremogram from the
residuals of the model fit can be used to check compatibility with the lack of extremal dependence.

\begin{figure}[ht]
\begin{center}
\centerline{\includegraphics[height=8cm,width=16cm]{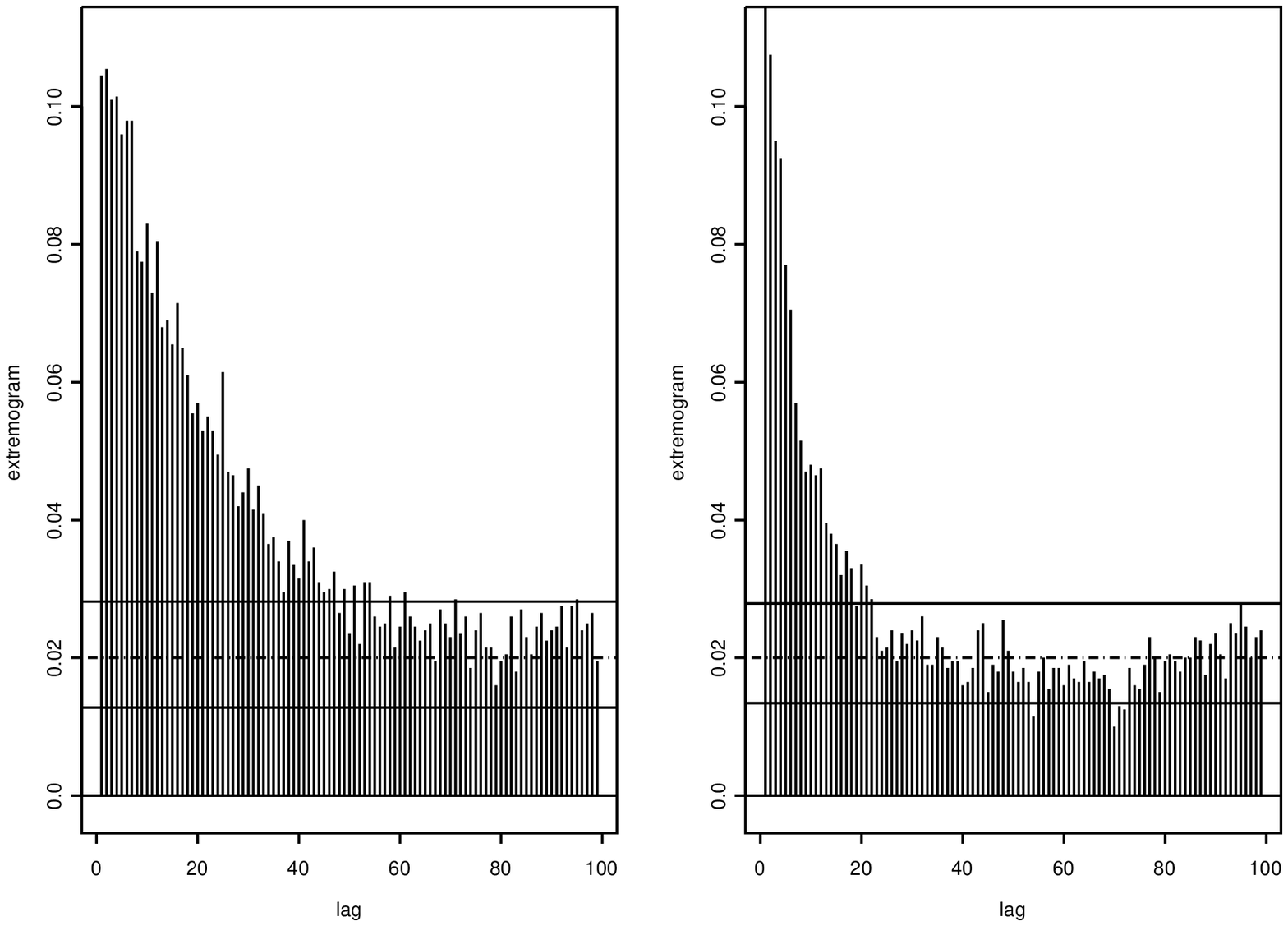}}
\end{center}
\bfi\label{fig:garchsv}{\small The sample extremogram for the upper tail for the \garch\ {\rm (left)} and SV processes {\rm
      (right)}, where the
    processes are specified in \eqref{eq:garch} and \eqref{eq:sv}, respectively. The sample size is $n=100,000$ and $a_m$ is the .98 empirical quantile. The solid horizontal lines are permutation-produced confidence bands and the dashed line at height .02 corresponds to the value of the PA-extremogram under independence.}
    \efi
\end{figure}

Figure~\ref{fig:garchsv}
shows the sample extremograms of a \garch\ (left) and a stochastic volatility  (right)
process. The GARCH realization was generated from the model,
\beam \label{eq:garch}
X_t=\sigma_tZ_t\,~~~~~~\mbox{and}~~~~~~\sigma^2_t= 0.1 +
0.14\,X_{t-1}^2+0.84\,\sigma_{t-1}^2\,,
\eeam
where  $(Z_t)$ is an iid sequence
with common distribution given by a $t_4$ (standardized to have variance 1).
The SV realization was produced from the model $(X_t)$ satisfying
\beam \label{eq:sv}
X_t=\sigma_tZ_t\,~~~~~~\mbox{and}~~~~~~\log \sigma_t = 0.9\,\log \sigma_{t-1}+\epsilon_t\,,
\eeam
where  $(\epsilon_t)$ is a sequence of iid  standard normal variables and is independent of the iid sequence $(Z_t)$, which has a $t_{2.6}$ distribution.
These conditions ensure that both the GARCH and SV realizations are regularly varying with index
$\alpha=2.6$, { i.e. they have power law tails with index $\alpha$; we
refer to Section~\ref{sec:ext} for a precise description.}
For the calculation of the sample extremograms, samples of size $n$ =
100,000 were used, the sets $A = B = (1,\infty)$ were chosen and, for
$a_m$, the $.98$ empirical quantile of the simulated data was taken.  It is evident from the slower decay of the sample extremogram for the
GARCH(1,1) process in Figure~\ref{fig:garchsv}
that this process exhibits extremal clustering while the faster
decay of the sample extremogram for the SV process
indicates the lack of clustering.
However, without a sense for the asymptotic distribution, it is virtually impossible to make any inferences about the extremogram (pre-asymptotic or otherwise).  Under the assumption of no serial dependence, one can compute permutation produced confidence bands (these are the solid lines in Figure \ref{fig:garchsv}).  Clearly, the extremogram for the GARCH is significantly greater than the dashed line at height .02, which corresponds to the value of the pre-asymptotic value of the extremogram under the null hypothesis of independence. On the other hand, the extremogram for the SV process tails off after lag 18 and is not significantly different than the .02 value that one would expect for independent data.  This is consistent with the theory described in Davis and Mikosch \cite{davis:mikosch:2009} in which there is extremal dependence for GARCH processes and none for SV processes.  The use of permutation procedures is illustrated in more detail in Section~\ref{sec:permutation}.

The remainder of the paper is organized as follows. A brief interlude into
the concept of regular variation on which the extremogram is built, is
provided in  Section~\ref{sec:ext}.  After establishing the theory of the bootstrapped extremogram in Section~\ref{sec:bs}, its use is demonstrated with several financial time series in Section~\ref{sec:bootext}. In conjunction with the bootstrapped extremogram, we present a quick and clean method for testing  {\it significant} serial extremal dependence using a random permutation procedure.  This procedure is actually similar in spirit to using the block bootstrap procedure, but with block size equal to 1.  The serial dependence is completely destroyed by randomly permuting the data so that the type I error for significance of the extremogram  under the null of no serial dependence can be controlled.

The  cross-extremogram for multivariate time series is defined and illustrated for real data examples including the returns of the major equity indices in Section~\ref{sec:further}.
Like the univariate time series, the cross-extremogram will depend on two sets, often decided upon the practitioner.
We use the
cross-extremogram to provide a method for assessing extremal
dependence between four major international stock markets, namely, the
FTSE 100, S\&P 500, DAX and Nikkei 225
Indices. Without controlling for the effect of changing volatility, the
cross-extremogram gives significant dependence between the series
for a large number of lags. However, after devolatizing each series
using a GARCH model, the resulting extremogram shows significant
dependence only at small lags. In addition, there is evidence of
directionality: large values of one index follow another
index.

In their presentation \cite{geman:chang:2009}, Geman and
Chang \cite{geman:chang:2009} consider the waiting times between rare
or extreme events for financial time series. They conclude from their
analysis that there is evidence of significant extremal clustering. In
Section~\ref{sec:Geman}, an extremogram that calculates these return
times between extreme events is defined and
the sample extremogram for various time series is provided as well. Consistent with the findings of Geman and Chang the presence of extremal clustering can be detected easily using the bootstrap.  The proof of the main theorems in Section~\ref{sec:bs} is provided in the Appendix.
\par
{ The reader who is mainly interested in applications of the
  sample extremogram to financial time series
  may skip the technical Sections~\ref{sec:ext} and \ref{sec:bs} and
  directly go to Sections~\ref{sec:bootext} and \ref{sec:further}.}

\section{Brief Interlude into Regular Variation}\label{sec:ext}\setcounter{equation}{0}
{ The extremogram \eqref{eq:ext} is a limit of conditional \pro ies and therefore
not always defined. In this section we give a sufficient condition for
its existence. The condition is rather technical; the interested
reader is referred
to Davis and Mikosch \cite{davis:mikosch:2009} for more details. We
also mention that this condition is satisfied for some of the standard
financial \ts\ models such as GARCH and SV; see
\cite{davis:mikosch:2001,davis:mikosch:2009,
davis:mikosch:2009a,davis:mikosch:2009b,davis:mikosch:2009c,lindner:2009}.
}

In this paper we focus on strictly stationary \seq s whose \fidi s
have power law tails in some generalized sense. In particular, we
will assume that the \fidi s of the $d$-dimensional process $(X_t)$
have {\em \regvary\ distributions with index $\alpha>0$.}
This means that for any $h\ge 1$, the radial part $|Y_h|$ of the lagged vector
$Y_h={\rm vec}(X_1,\ldots,X_h)$ is {\em regularly varying with tail
  index $-\alpha$}\ :
\beao
\dfrac{P(|Y_h|>t\,x)}{P(|Y_h|>x)} \rightarrow t^{-\alpha}\quad \mbox{as
$\xto$,}\quad t>0\,,
\eeao
and the angular part $Y_h/|Y_h|$ is asymptotically
independent of the radial part $|Y_h|$ for large values of $|Y_h|$:
for every $h\ge 1$, there exists a random vector $\Theta_h \in \bbs^{hd-1}$ such that
\beao
P(Y_h/|Y_h| \in \cdot \mid |Y_h|>x)\stw P(\Theta_h \in
\cdot)\quad\mbox{as $\xto$}.
\eeao
Here $\stw$ denotes weak convergence on the Borel $\sigma$-field of
$\bbs^{hd-1}$, the unit sphere in $\bbr^{hd}$ with respect to a given
norm $|\cdot|$.
The \ds\ $P(\Theta_h \in \cdot)$ is called the {\em spectral measure} and
$\alpha$ the {\em index of the \regvary\ vector $Y_h$.} We also refer to $(X_t)$ as a
{\em \regvary\ \seq\ with index $\alpha$.}
\par
For our purposes it will be convenient to use a sequential definition
of a \regvary\ \seq\ $(X_t)$ which is equivalent to the definition above:
there exists a \seq\ $a_n\to \infty$, an $\alpha>0$ and a \seq\ of
non-null Radon \ms s $(\mu_h)$ on the Borel $\sigma$-field of
$\ov \bbr_0^{hd}=\ov \bbr^{hd}\backslash \{\bf0\}$ \st\ for $h\ge 1$,
\beam\label{eq:regvar3}
n\,P(a_n^{-1}Y_h \in \cdot ) \stv \mu_h(\cdot)\,,
\eeam
where $\stv $ denotes vague \con\ on the same $\sigma$-field; see Resnick \cite{resnick:2007},
Section~6.1. The limiting \ms s have the property
$\mu_h(t\cdot)= t^{-\alpha}\mu_h(\cdot)$, $t>0$. We refer to Basrak and Segers
\cite{basrak:segers:2009} who give an enlightening interpretation of the
structure of a \regvary\ \seq .

Now to connect the measure in \eqref{eq:regvar3} with the extremogram, for suitably chosen sets $A$ and $B$ in $\bbr^d$ bounded away from the origin, set for $h\ge 2$, $\tilde A=A\times \bbr^{(h-1)d}$ and $\tilde B=A\times \bbr^{(h-2)d}\times B$.  Provided $\tilde A$ and $\tilde B$ are $\mu_h$ continuity sets, then
\beam \label{eq:rhomu}
\rho_{A,B}(h-1)=\lim_{n\to\infty}P(a_n^{-1}X_{h}\in B\mid  a_n^{-1}X_1\in A)
=\lim_{n\to\infty}\frac{nP(a_n^{-1}Y_h \in \tilde B)}{nP(a_n^{-1}Y_h \in \tilde A)}
=\frac{\mu_h(\tilde B)}{\mu_h(\tilde A)}\,.
\eeam
It is worth noting that standard arguments in regular variation allow one to replace $a_n$ in the first limit appearing in \eqref{eq:rhomu} with any sequence of numbers $x_n$ tending to $\infty$.

\section{The Bootstrapped Sample Extremogram}\label{sec:bs}\setcounter{equation}{0}
 In this section we will construct confidence bands for the sample
extremogram based on a bootstrap procedure
which takes into account the serial dependence structure of the data.
The resulting confidence bands closely follow the sample
extremogram. Moreover, assuming \regvar\ of the underlying \ts , we
will be able to show that the bootstrap confidence bands are asymptotically correct.

\subsection{Stationary bootstrap}\label{subsec:bs}
This resampling scheme was introduced
by Politis and Romano \cite{politis:romano:1994}. It is an adaptation
of the block bootstrap which allows for randomly varying
block sizes.
\par
For any strictly stationary \seq\ $(Y_t)$
the {\em stationary bootstrap procedure} consists of generating
pseudo-samples $Y_1^\ast,\ldots,Y_n^\ast$ from the sample
$Y_1,\ldots,Y_n$ by taking the first $n$ elements from
\beam\label{eq.sample}
Y_{K_1},\ldots,Y_{K_1+L_1-1},\ldots,Y_{K_N},\ldots,Y_{K_N+L_N-1}\,,
\eeam
where $(K_i)$ is an iid \seq\ of \rv s uniformly distributed
on $\{1,\ldots,n\}$, $(L_i)$ is an iid \seq\ of geometrically
distributed random variables with \ds\
$P(L_1=k)=p\,(1-p)^{k-1}$, $k=1,2,\ldots,$ for some $p=p_n\in
(0,1)$ \st\ $p_n\to 0$ as $\nto$, and
\beao
N=N_n=\inf\{i\ge 1:L_1+\cdots
+L_i\ge n\}\,.
\eeao
The upper limits of the random blocks $\{K_i,\ldots,K_i+L_i-1\}$
may exceed the sample size $n$. Therefore, in \eqref{eq.sample} we
replace the (unobserved) $Y_t$'s with $t>n$
by the observations $Y_{t\;{\rm mod}\; n}$.
Finally, the three \seq s
$(Y_t)$, $(K_i)$ and $(L_i)$ are also supposed independent.
The dependence of the \seq s $(K_i)$ and $(L_i)$ on $n$ is
suppressed in the notation. The generated pseudo-sample
can be extended to an infinite \seq\ $(Y_t^\ast)$
by extending \eqref{eq.sample} to an infinite \seq .
For every
fixed $n\ge 1$, $(Y_t^\ast)$ constitutes a strictly stationary \seq .

\subsection{Main results}
We want to apply the stationary bootstrap procedure to the
strictly stationary \seq\ of the indicator \fct s $I_t= I_{\{a_m^{-1}X_t\in C\}}$,
$t\in \bbz$, where the underlying \seq\ $(X_t)$ is strictly
stationary $\bbr^d$-valued and \regvary\ with index $\alpha$,
$a_m$ has the interpretation as a high quantile of the \ds\ of
$|X_0|$,
and $C$ is a set bounded away from zero.
The application of the stationary bootstrap in this context is rather
unconventional since the \seq s $(I_t)$ constitute triangular arrays of strictly
stationary \seq s: through $a_m$ these \seq s also depend on $n$.
\par
We write $(I_t^\ast)$ for a bootstrap \seq\ generated from the sample
$I_1,\ldots,I_n$ by the stationary bootstrap procedure described
above. In what follows, $P^\ast$, $E^\ast$ and $\var^\ast$ denote
the \pro y \ms\ generated by the bootstrap procedure, the
corresponding expected value and variance. This means that
$P^\ast(\cdot)=P(\cdot \mid (X_t))$ is the infinite
product \ms\ generated by the \ds s of $(K_i)$ and $(L_i)$.
\par
The bootstrap sample mean
$\ov I_n^\ast=n^{-1} \sum_{i=1}^n I_i^\ast$
satisfies the following elementary properties:
\beao
E^\ast (\ov I_n^\ast)&=&E^\ast (I_1^\ast)= \ov I_n= n^{-1} \sum_{i=1}^n I_i\,,\\[2mm]
s_n^2=\var^\ast(n^{1/2}\ov I_n^\ast)&=& C_n(0)+ 2\sum_{h=1}^{n-1}
(1-h/n)\,(1-p)^h\,C_n(h)\,,
\eeao
where
\beao
C_n(h)=n^{-1} \sum_{i=1}^n (I_i-\ov I_n)(I_{i+h}-\ov I_n)\,,\quad
h=0,\ldots,n\,,
\eeao
are the {\em circular sample autocovariances.} Here we again made use
of the circular construction $I_j=I_{j \,{\rm mod}\,n}$.
Writing
\beao
\gamma_n(h)=n^{-1}\sum_{i=1}^{n-h} (I_i-\ov I_n)(I_{i+h}-\ov
I_n)\,,\quad h\ge 0\,,
\eeao
for the ordinary {\em sample autocovariances,} we have
\beam\label{eq:5}
C_n(h)= \gamma_n(h)+\gamma_n(n-h)\,,\quad h=0,\ldots,n\,.
\eeam
\par
Recall the notion of a strictly stationary \regvary\ \seq\ from
Section~\ref{sec:ext}, in particular the \seq\ of limiting \ms s
$(\mu_{h})$; see \eqref{eq:regvar3}. For convenience, we write
$\mu=\mu_1$.
For any subset $C\subset \ov \bbr_0^{d}$, define the quantities
\beam
\sigma^2(C)=\mu(C)+ 2\,
\sum_{h=1}^\infty \tau_h(C)\,,\label{eq:2}
\eeam
with $\tau_h(C)=\mu_{h+1}(C\times \ov \bbr^{d(h-1)}_0\times C)$
and, suppressing the dependence on $C$ in the notation,
\beao
\wh P_m&=& m \,\ov I_n\,,\quad p_0=EI_1 =P(a_m^{-1}X_1\in C)\,,\\\quad
p_{0h}&=&E(I_0I_{h})=P(a_m^{-1}X_0\in C\,,a_m^{-1}X_h\in C)\,,\quad
h\ge 1\,.
\eeao
We also need the following mixing condition:
\begin{enumerate}
\item[\rm (M)] The \seq\ $(X_n)$ is strongly mixing with rate \fct\
  $(\alpha_t)$. Moreover, there exist $m=m_n\to\infty$ and
  $r_n\to\infty$
\st\ $m_n/n\to 0$ and $r_n/m_n\to 0$ and
\beam\label{eq:7}
\lim_{\nto} m_n\sum_{h=r_n}^\infty \alpha_h=0\,,
\eeam
and for all $\epsilon>0$,
\beam\label{eq:6}
\lim_{k\to\infty}\limsup_{\nto} m_n\sum_{h=k}^\infty P(|X_h|>\epsilon
\,a_m\,,|X_0|>\epsilon\, a_m)=0\,.
\eeam
\end{enumerate}
\par
Our next goal is to show that the stationary bootstrap is
\asy ally correct for the bootstrapped estimator of $\wh P_m$ given by
\beao
\wh P_m^\ast= m\,\ov I_n^\ast= \dfrac{m}{n} \sum_{t=1}^n I_t^\ast\,.
\eeao
The following result is the stationary bootstrap analog of
Theorem~3.1 in \cite{davis:mikosch:2009}. It shows that the bootstrap estimator $\wh P_m^\ast$
of $\wh P_m$ is \asy ally correct.
\bth\label{thm:1}
Assume that the following conditions hold for the strictly stationary
\regvary\ \seq\ $(X_t)$ of $\bbr^d$-valued random vectors:
\begin{enumerate}
\item
The mixing condition {\rm (M)} and in addition
\beam\label{eq:10}
\sum_{h=1}^\infty k\, \alpha_k<\infty\,.
\eeam
\item The growth conditions
\beam\label{eq:9}
p=p_n\to 0\,,\quad\mbox{and}\quad n\,p^2/m\to\infty.
\eeam
\item
The sets $C$ and $C\times \ov
\bbr_0^{d(h-1)}\times C\subset \ov \bbr_0^{d(h+1)}$
are continuity sets \wrt\ $\mu$ and $\mu_{h+1}$ for $h\ge 1$,
$C$ is bounded away from zero and $\sigma^2(C)>0$.
\item
The central limit theorem, $\left(n/m\right)^{1/2}(\hat P_m-mp_0)\cid N(0,\sigma^2(C))$ holds.
\end{enumerate}
Then the following bootstrap consistency results hold:
\beam\label{eq:1}
E^\ast (\wh P_m^\ast) &\stp& \mu(C)\,,\\[2mm]
m\,s_n^{2}=\var^\ast((n/m)^{1/2}\wh P_m^\ast)&\stp& \sigma^2(C)\,,\label{eq:2a}
\eeam
with $\sigma^2(C)$ given in \eqref{eq:2}.  In particular
\beam\label{e:hilfg}
P^\ast(|\wh
P^\ast_m-\mu(C)|>\delta) \stp 0\,,\quad \delta>0\,,
\eeam
and the \clt\ holds
\beam\label{eq:3}
\sup_x\left|P^\ast\big((n/m)^{1/2}(ms_n^2)^{-1/2} (\wh P_m^\ast
      -\wh P_m)\le x \big)
-\Phi(x)\right|&\stp& 0\,,
\eeam
where $\Phi$ denotes the standard normal \df.
\ethe
Politis and Romano \cite{politis:romano:1994} proved a corresponding
result for the sample mean of the stationary bootstrap \seq\
$(X_t^\ast)$ for a finite variance strictly stationary \seq\
$(X_t)$. They also needed the growth conditions $p_n\to 0$ and
$n\,p_n\to \infty$. Our additional condition $(n\,p_n)
(p_n/m)\to\infty$, which implies $n\,p_n \to \infty$ is needed
since $\wh P_m^\ast$ is an average in the
triangular scheme $I_t=I_{\{a_m^{-1}X_t\in C \}}$, $t=1,\ldots,n$.
Although various steps in the proof are similar to those in
Politis and Romano \cite{politis:romano:1994}, the triangular nature of
the bootstrapped \seq\ $(I_t)$ requires some new ideas. We found it
surprising that the full stationary bootstrap works in this
context. In the context of extreme value statistics the
bootstrap often needs to be modified even when the data are iid.
\par
We now turn our attention to the sample extremograms for which both the numerator and denominator
are estimators of the type $\wh P_m$. Therefore our next objective are
the \asy\ properties of these ratio estimators which we study in a
general context.
We consider general sets $D_1,\ldots, D_{h}\subset \ov \bbr_0^d$
and $C=D_{h+1}\subset \ov \bbr_0^d$, $h\ge 1$. Since we deal with
several sets $D_i$
we need to indicate that the indicator \fct s $I_t$ depend on these sets:
\beao
I_t(D_i)= I_{\{a_m^{-1}X_t\in D_i\}}\,,\quad t\in\bbz \,,
\eeao
and we proceed similarly for the estimators $\wh P_m(D_i)$ of
$\mu(D_i)$.
Now we define the corresponding {\em ratio estimators}
\beao
\wh\rho _{C,D_i}= \dfrac{\wh P_m(D_i)}{\wh
  P_m(C)}=\dfrac{\sum_{t=1}^n I_{\{a_m^{-1}X_t\in D_i\}}}{\sum_{t=1}^n I_{\{a_m^{-1}X_t\in C\}}}\,,\quad i=1,\ldots,h\,.
\eeao
Davis and Mikosch \cite{davis:mikosch:2009}, Corollary 3.3, proved the joint \asy\
normality of these estimators:
Note that there is a misprint for the expression for $r_{D_i,D_j}$ in \cite{davis:mikosch:2009}, which we now
correct here as
\beao
r_{D_i,D_j}&=& \mu(D_i\cap D_j)+\sum_{h=1}^\infty[\mu_{h+1}(D_j\times \ov\bbr_0^{d(h-2)}\times D_i)+
\mu_{h+1}(D_i\times \ov\bbr_0^{d(h-2)}\times D_j]\,.
\eeao
The centering in the \clt\ of Corollary 3.3 of \cite{davis:mikosch:2009} uses the PA-extremogram as opposed to the extremogram.  In general the PA-extremograms cannot be replaced by their limits
\beao
\rho_{C,D_i}=\dfrac{\mu(D_i)}{\mu(C)}\,, \quad i=1,\ldots,h\,,
\eeao
unless the following additional condition holds
\beam\label{eq:pp9}
\lim_{\nto} \sqrt{nm_n}\, [\mu(D_i) P(a_m^{-1}X_0\in
C)-\mu(C)\,P(a_m^{-1}
X_0\in D_i)]=0\,,\quad i=1,\ldots,h\,,
\eeam
In addition to the complex form of the \asy\ variance
which can hardly be evaluated,
condition \eqref{eq:pp9} points at another practical problem when applying
the \clt\ to the ratio estimators. The next result will show that
these problems will be overcome by an application of the stationary bootstrap.
\par
We construct bootstrap samples $I_1^\ast(D_i),\ldots,I_n^\ast(D_i)$,
$i=1,\ldots,h+1$, from the samples
$I_1(D_i),\ldots, $ $I_n(D_i)$, $i=1,\ldots,h+1$, by a simultaneous application of the
stationary bootstrap procedure, i.e., we use the same \seq s $(K_i)$
and $(L_i)$ for the construction of the $h+1$ bootstrap samples; see Section~\ref{subsec:bs}.
From the bootstrap samples the
bootstrap versions $\wh P_m^ \ast (D_i)$ of
$\wh P_m(D_i)$, $i=1,\ldots,h+1$,
and $\wh{\rho}^\ast_{C,D_i}$ of $\wh{\rho}_{C,D_i}$, $i=1,\ldots,h$,
are constructed.
\par
The following result shows that the bootstrapped ratio estimators $\wh
\rho_{C,D_i}^\ast$ are
\asy ally correct estimators of their sample counterparts $\wh
\rho_{C,D_i}$, $i=1,\ldots,h$.
\bth\label{thm:2}
Assume that the following conditions hold for the strictly stationary
\regvary\ \seq\ $(X_t)$ of $\bbr^d$-valued random vectors:
\begin{enumerate}
\item
The mixing conditions {\rm (M)} and
\eqref{eq:10} hold.
\item The growth conditions \eqref{eq:9} on $p=p_n$ hold.
\item
The sets $D_1,\ldots, D_{h+1}$,
$D_i\times \ov \bbr_0^{d(i-1)}\times D_i \subset \ov \bbr_0^{(i+1)d}$
are continuous \wrt\ $\mu$ and $\mu_{i+1}$, $i=1,\ldots,h+1$,
$\mu(C)>0$ and $\sigma^2(D_i)>0$, $i=1,\ldots,h$.
\item
The \clt\ $(n/m)^{1/2}\left(\widehat\rho_{C,D_i}-\rho_{C,D_i:m}\right)_{i=1,\ldots,h}\cid N({\bf0}, \Sigma)$ holds, where the asymptotic variance is defined in Corollary 3.3 of \cite{davis:mikosch:2009}.
\end{enumerate}
Then the bootstrapped ratio estimators satisfy the following
bootstrap consistency result
\beam\label{eq:hilfb}
P^\ast\big( |\wh \rho_{C,D_i}^\ast-\rho_{C,D_i}|>\delta\big)\stp
0\,,\quad \delta>0\,,
\eeam
and the \clt\ holds
\beam\label{eq:cltbs}
P^\ast( (n/m)^{1/2} \big( \wh\rho_{C,D_i} ^\ast-\wh
\rho_{C,D_i}\big)_{i=1,\ldots,h}\in A) \stp \Phi_{{\bf0},
    \Sigma}(A)\,,
\eeam
where $A$ is any continuity set of the normal \ds\
$\Phi_{{\bf0},\Sigma}$
with mean zero and covariance matrix~$\Sigma$.
\ethe

\subsection{Consistency for the bootstrapped sample extremogram}\label{subsec:consist}
Recall the definition of the sample extremogram $(\wh\rho_{A,B}(i))$
from \eqref{eq:sext}.
This estimate can be recast as a ratio estimator by introducing the
$\bbr^{d(h+1)}$-valued vector process
\beam\label{eq:stick}
Y_t={\rm vec}(X_t,\ldots,X_{t+h}),\quad t\in \bbz\,,
\eeam
consisting of stacking $h+1$ consecutive values of the time series
$(X_t)$.
Now the sets $C$ and $D_0,\ldots,D_h$ specified in
Theorem~\ref{thm:2}
are defined through the relations $C=A\times \ov\bbr_0^{dh}$,
$D_0=A\cap B\times \ov\bbr_0^{dh}$ and for
$D_i=A\times \ov\bbr_0^{d(i-1)}\times B\times \ov\bbr_0^{d(h-i)}$ for
$i\ge1$. With this convention, Theorem~\ref{thm:2}
can be applied to the $(Y_t)$ and $(Y_t^\ast)$ sequences directly. We formulate here
the consistency result for the bootstrapped sample extremogram
\beao
\wh\rho_{AB}^\ast(i)=
\dfrac{
\sum_{t=1}^{n-i} I_{\{ a_m^{-1}X_t^\ast
    \in A,a_m^{-1} X_{t+i}^\ast\in B\}}}
{\sum_{t=1}^{n} I_{\{ a_m^{-1}X_t^\ast
    \in A\}}}\,,\quad i\ge 0\,.
\eeao
We use the same notation as in Theorem~\ref{thm:2}.
\bco \label{cor:1}
Assume that the conditions of Theorem~\ref{thm:2}
are satisfied for the \seq\ $(Y_t)$ and the sets $C,D_0,\ldots,D_{h}$
defined above.
Then, conditionally on $(X_t)$,
\beao
(n/m)^{1/2}\,\big(
\wh\rho_{AB}^\ast(i)-
    \wh\rho_{AB}(i)\big)_{i=0,1,\ldots,h}\std
  N({\bf0}, { \Sigma})\,.
\eeao
\eco
\par
For later use, we will mention a result for the
{\em return times extremogram} $(\rho_A(i))$ given by the limit relations
\eqref{eq:rhotime} for a fixed set $A\subset \ov \bbr_0^d$ bounded
away from zero
and the corresponding {\em return times sample extremogram}
$(\wh\rho_{A}(i))$
defined in \eqref{eq:GCext}. The bootstrapped return times sample
extremogram
$(\wh\rho_{A}^\ast(i))$ is defined in the straightforward way by
replacing $(X_t)$ by the stationary bootstrap \seq\ $(X_t^\ast)$.
We again use the vector process $(Y_t)$ defined in \eqref{eq:stick}
and sets $C=A\times \ov\bbr_0^{dh}$ and $D_i=A\times (A^c)^{i-1}\times
A\times \ov\bbr_0^{d(h-i}$, $i=1,\ldots,h$, to recast
$(\wh\rho_{A}(i))_{i=1,\ldots,h}$ and $(\wh\rho_{A}^\ast
(i))_{i=1,\ldots,h}$, $h\ge 1$, as ratio estimators.
Then Theorem~\ref{thm:2} yields \clt s for the corresponding sample extremogram
and its bootstrap version which we omit. These results show that the
stationary bootstrap is \asy ally correct for the considered return
times extremogram.

\section{Examples of the Bootstrapped Sample Extremogram}\label{sec:bootext}\setcounter{equation}{0}

\begin{figure}[ht]
\begin{center}
\centerline{\includegraphics[height=8cm,width=16cm]{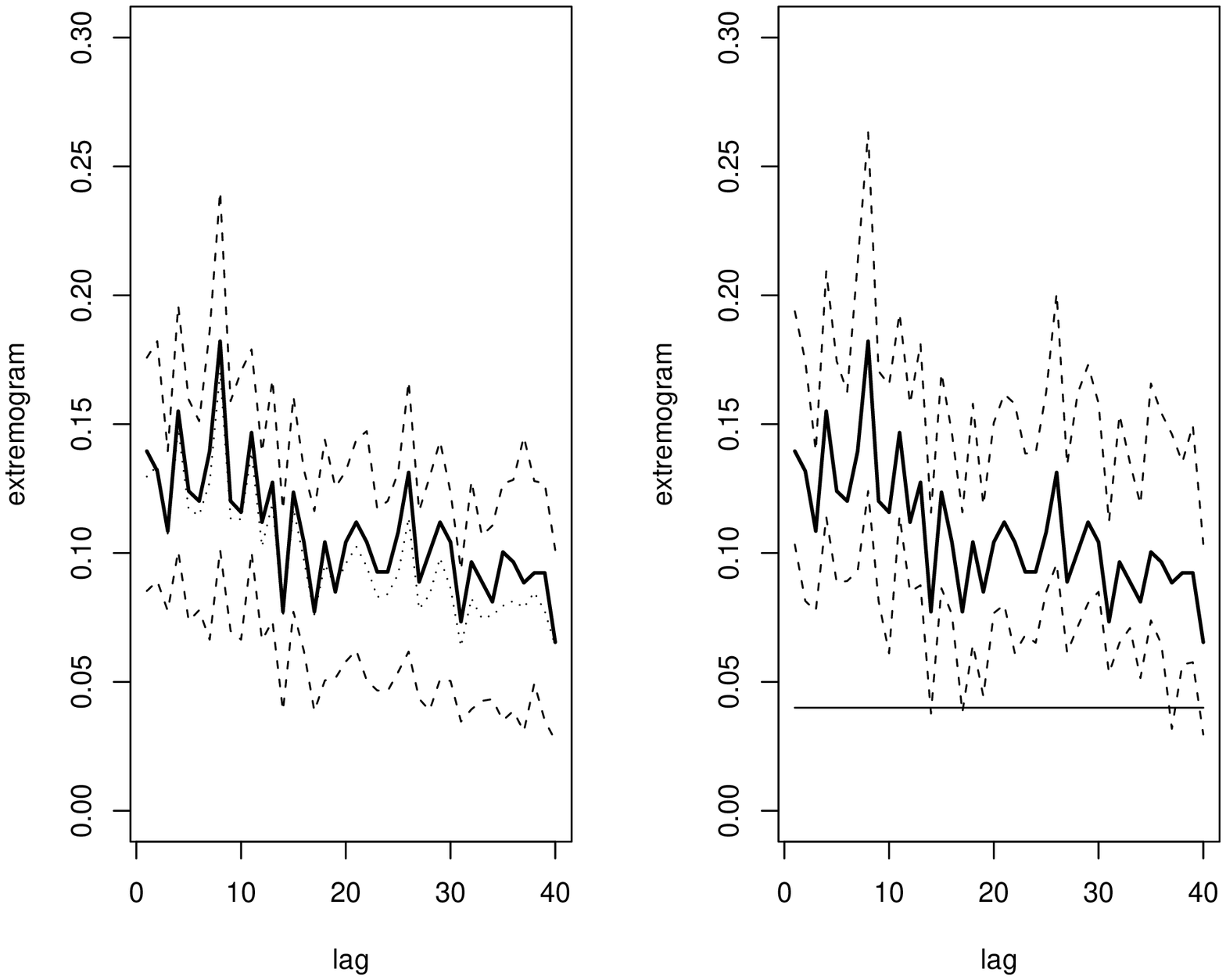}}
\end{center}
\bfi{\small
{\rm Left: } The .975 and .025 empirical quantiles (dashed lines) of 10,000 bootstrapped replicates of the
extremogram for the daily log-returns of the FTSE; the sample extremogram (solid line) and the mean of the bootstrapped replicates (dotted line).
{\rm Right:}
95\% confidence bands (dashed lines) for the PA-extremogram based on a bootstrap approximation to the sampling distribution to the sample extremogram (solid line) and the PA-extremogram (horizontal line at .04) based on the data being independent.}
\label{fig:booteg}\efi
\end{figure}

The first application of the bootstrapped extremogram is to the 6,443 daily log-returns of the FTSE 100 exchange from April 4, 1984 to October 2, 2009.
The sample extremogram of the FTSE for lags 1 to 40 corresponding
to the left tail ($A=B=(-\infty,-1)$) with $a_m$ equal to the negative of the $.04$ empirical quantile is displayed as the bold lines in both panels of Figure ~\ref{fig:booteg}.  In the left graph, the dashed lines represent .975 and .025 quantiles of
the sampling distribution of $\wh\rho_{A,B}^{*}(h)$
based on 10,000 bootstrap replications for the daily
log-returns of the FTSE.  The dotted line is the mean of the bootstrapped replicates.  The bootstrapped extremogram
decays slowly to zero which signifies extreme serial dependence.
\par
The right graph in Figure ~\ref{fig:booteg}
shows approximate 95\% confidence intervals (dashed lines) for the PA-extremogram that are found using the appropriate cutoff values from the empirical distribution of the bootstrapped replicates of $\hat\rho_{A,B}^*(h)-\hat\rho_{A,B}(h)$ and the sample extremogram (dark solid line).  Notice that due to the bias in the bootstrapped distribution, the sample extremogram does not fall in the center of the intervals.  Using a small $p_n$ in the bootstrapped replicates helps reduce this bias. Observe that the horizontal solid line at height .04, corresponding to a PA-extremogram under an independence assumption, is well outside these confidence bands confirming the serial extremal dependence.

\begin{figure}[ht]
\begin{center}
\centerline{\includegraphics[height=8cm,width=16cm]{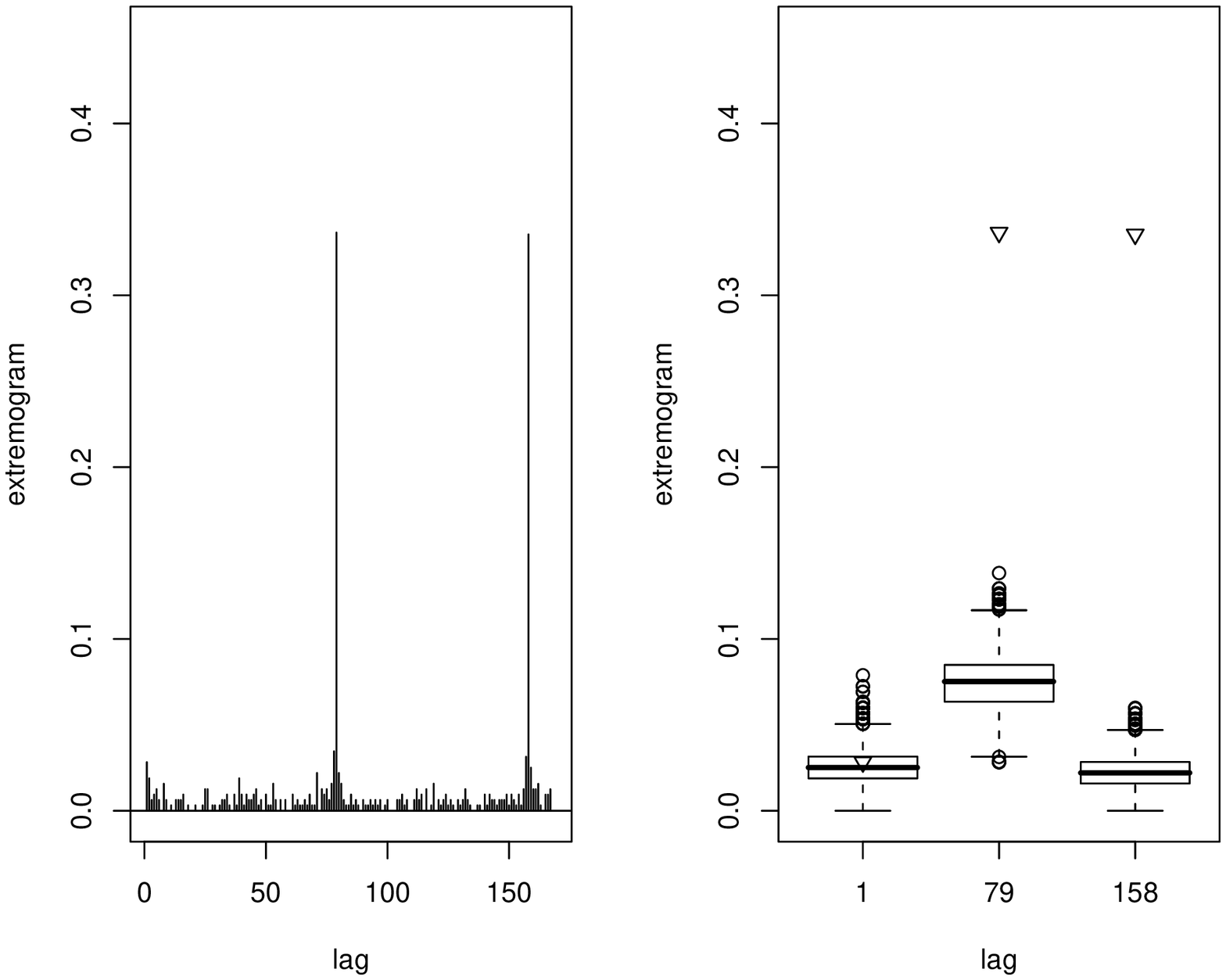}}
\end{center}
\begin{center}
\centerline{\includegraphics[height=8cm,width=16cm]{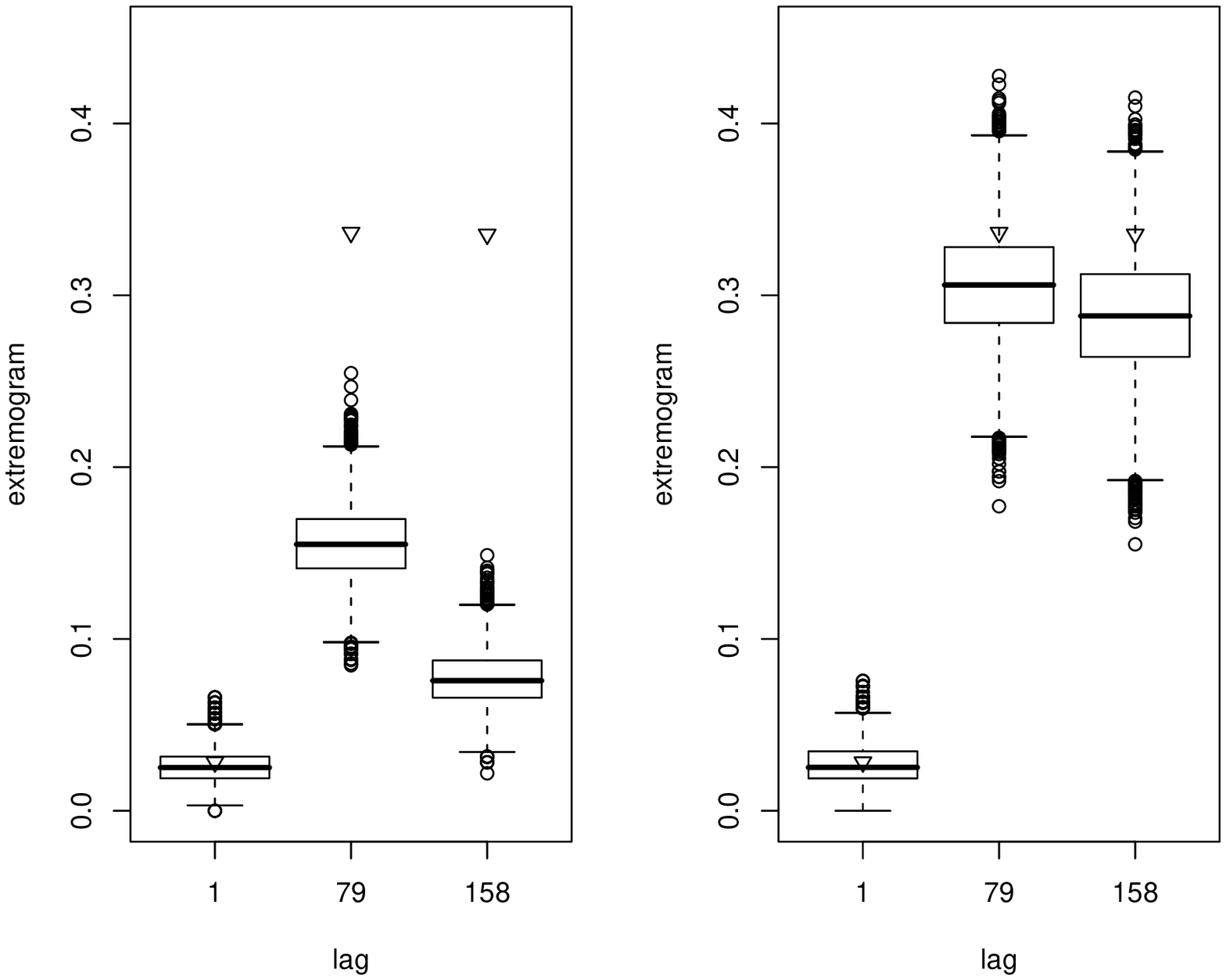}}
\end{center}
\bfi{\small The sample extremogram for the $5$-minute
log-returns of GS {\rm (top left)}.
Boxplots of the corresponding bootstrapped replicates of the extremograms at lags 1, 79 and 158, using a mean
block size of $50$ {\rm (top right)}, $100$ {\rm (bottom left)}
and $200$ {\rm (bottom right)}.}
\label{fig:bootgs}
\efi
\end{figure}

The next example illustrates how the stationary bootstrap for
the sample extremogram performs as a function on the choice of the mean
block size.  Recall from Section~\ref{sec:bs} that the condition $p_n\to 0$
is needed in order to achieve consistency of the bootstrap
estimators of the extremogram. Figure ~\ref{fig:bootgs} (top left)
shows the sample extremogram of the left tail for the 5-minute log-returns of
Goldman Sachs (GS) from December 1, 2004 to July 26, 2006.
We choose sets $A = B = (-\infty,-1)$ and
the negative of the $.01$ empirical quantile of the log-returns for $a_m$. The
remaining graphs in
Figure ~\ref{fig:bootgs} show the boxplots of the sampling distribution for bootstrap replicates of the sample extremogram at just the most interesting lags of 1, 79, and 158.


The sample extremogram has a large spike at
lags 79 and 158. The New York Stock Exchange (NYSE) is open
daily from 9:30am to 4pm. Hence, there are 78 5-minute
spells each day and so we can conclude that there
is evidence of strong extremal dependence between returns a day apart.
In the top right graph ($p_n=1/50$) the .975 quantiles of the distribution of the bootstrap
at lags 79 and 158
do not reach the same height as in the bottom right
graph ($p_n=1/200$).
By resampling blocks with mean block size 50, the assumption is made that
the dependence in observations $X_{t}$ and $X_{t+k}$ for $k>50$ has little, if any, impact on the distribution of the sample extremogram.
In particular, the dependence
structure is broken for lags greater than 50 and so the bootstrapped
sample extremogram cannot
capture the extent of the extremal dependence at lag 79 and beyond,
and certainly not at lag 158.
\par
The fixed block bootstrap is a method that resamples
blocks of data of fixed length in order to keep the dependence
structure intact. If such a bootstrap with fixed block size 50
was used, the opportunity to detect the extremal dependence
at lags 79 and 158 would be impossible. Thus, the randomly chosen
block size is a potential advantage of the stationary bootstrap:
it is not unlikely to get blocks of size 90, say, if the mean block size is
50. As the mean block sizes increase from 50 to 100 to 200,
the stationary bootstrap captures more and more of
the dependence at lags 79 and 158.

\subsection{A random permutation procedure.}\label{sec:permutation}
We use a permutation test procedure to produce {\it confidence bands}  for the sample extremogram under the assumption that the underlying data are in fact independent.  These bounds can be viewed as the analogue of the standard $\pm 1.96/\sqrt{n}$ bounds used for the sample autocorrelation function (ACF) of a time series.  Values of the sample ACF that extend beyond these bounds lend support that the ACF at the corresponding lags are non-zero.  The bounds for the sample ACF are based on well-known asymptotic theory.  Unfortunately, such bounds are not easily computable for the sample extremogram $\hat\rho_{A,B}(h)$.  For a fixed lag $h$, if the value of the sample extremogram for the original data is not extreme relative to the values of the sample extremogram based on random permutations of the data, then the sample extremogram is impervious to the time order of the data.  On the other hand, if the $\hat\rho_{A,B}(h)$ is more extreme (either larger or smaller than all the extremograms computed for 99 random permutations of the data), then we conclude the presence of extremal dependence at lag $h$ with probability .98=98/100.  Aside from boundary effects, (i.e., the numerator of the sample extremogram is a sum over $n-h$ terms and hence depends mildly on $h$), the permutation distribution of the sample extremogram is virtually the same for all lags $h$.  The bold lines in the graphs of Figure ~\ref{fig:garchsv} correspond to the maximum and minimum of the sample extremogram at lag 1 based on 99 random permutations of both the GARCH (left panel) and the SV (right panel) models.  The dashed line is the value of the PA-extremogram under the assumption that the data are in fact independent.  In this case the value is .02.  Clearly, the extremogram is measuring extremal dependence in both series.

\subsection{Equity indices.}\label{sec:EQIndex}
\begin{figure}[ht]
\begin{center}
\centerline{\includegraphics[height=8cm,width=16cm]{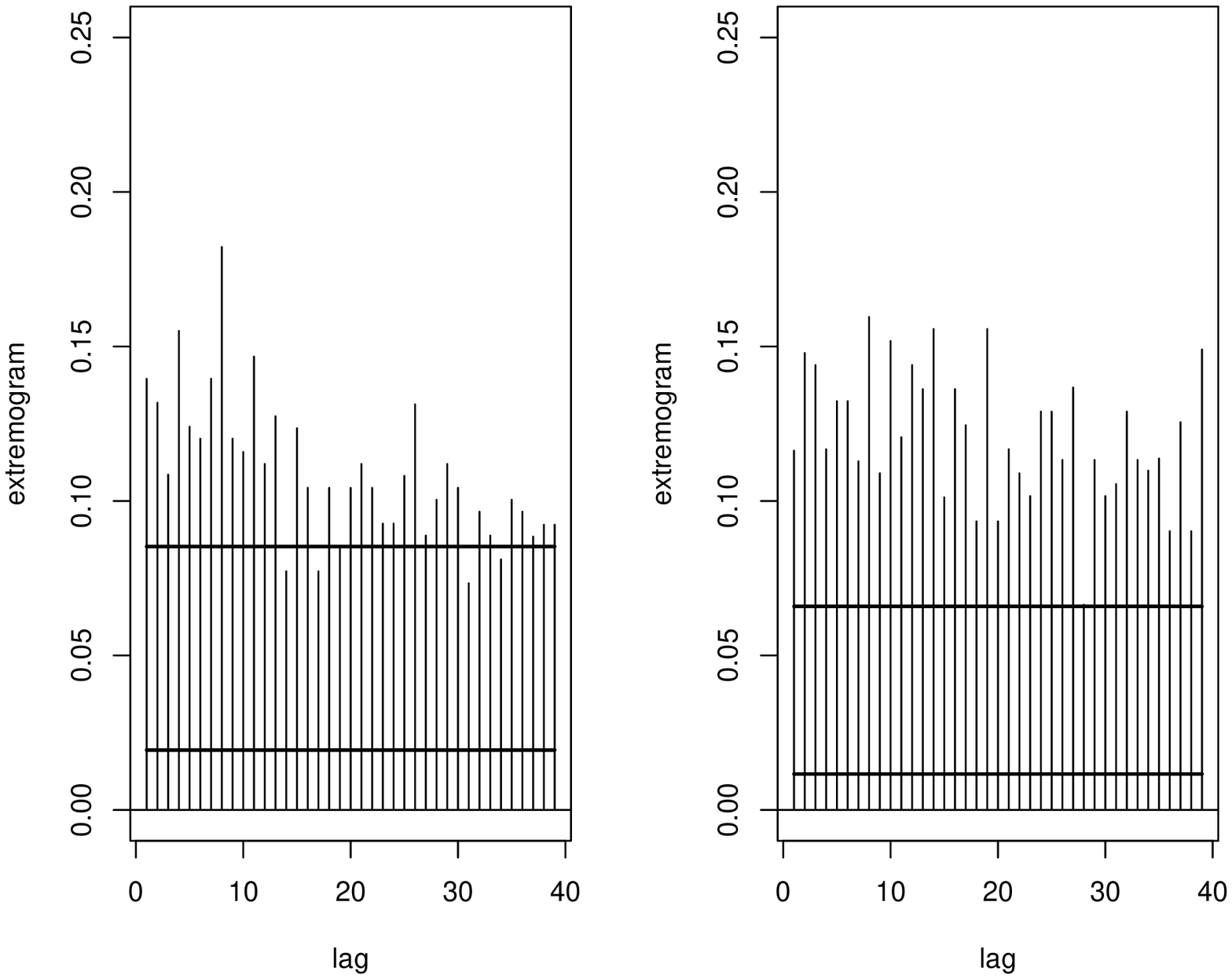}}
\end{center}
\begin{center}
\centerline{\includegraphics[height=8cm,width=16cm]{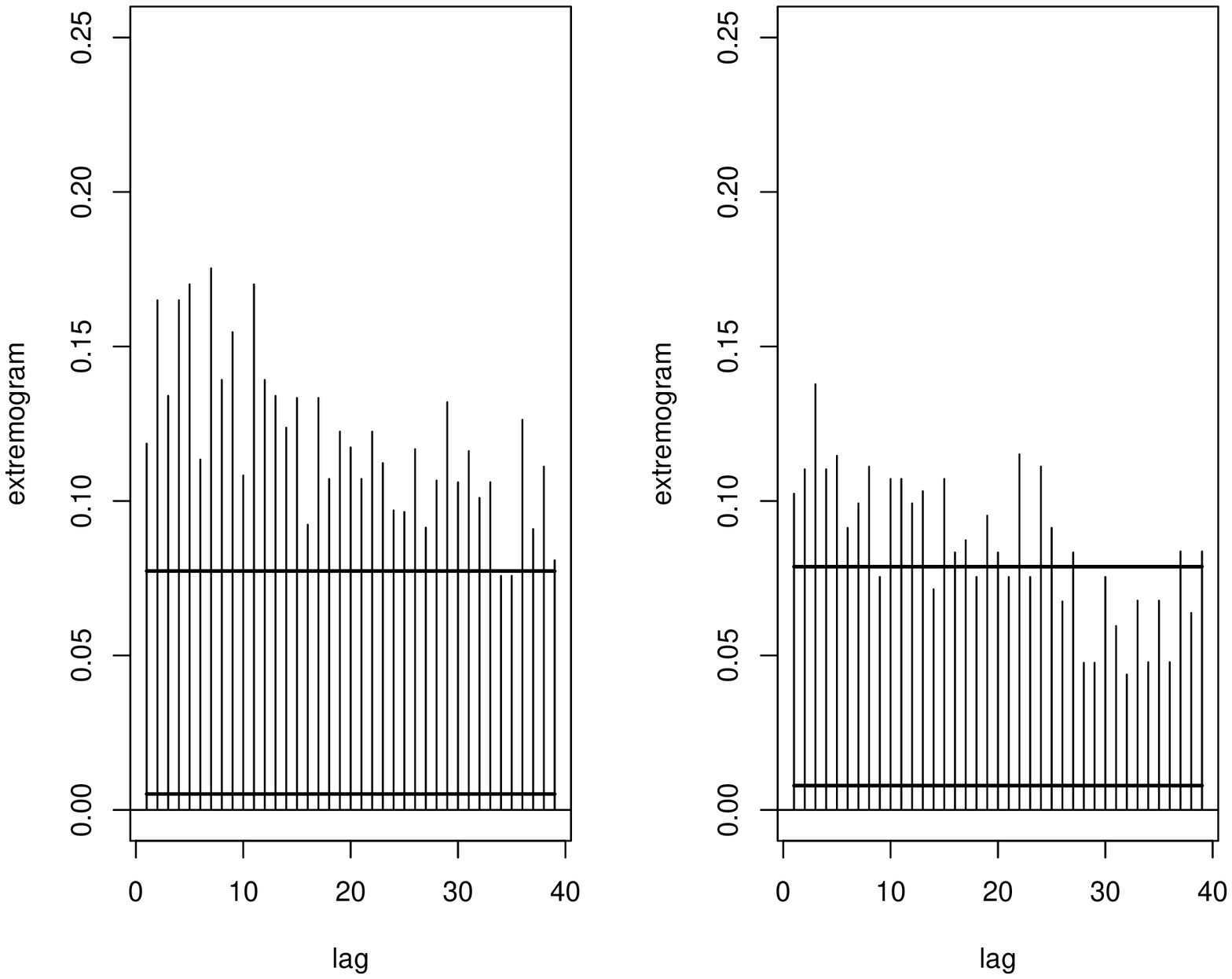}}
\end{center}
\bfi{\small The sample extremogram for the lower tails of the FTSE {\rm (top left)}, S\&P
  {\rm (top right)}, DAX {\rm (bottom left)} and Nikkei. The bold
  lines represent sample extremograms based on $99$ random permutations of the data.}\label{fig:FTSESP}
\efi
\end{figure}

For the next example we consider daily equity index log-returns for
four countries: the United States, the United Kingdom, Germany and
Japan. Here and in what follows, the indices are left in their local currencies. The top left and right graphs in
Figure ~\ref{fig:FTSESP}
show the sample extremograms for the negative tails ($A=B=(-\infty,-1]$ with $a_m$ estimated as the absolute value of the .04 empirical quantile) applied to 6,443 daily
log-returns of the FTSE 100 and S\&P 500 Indices from April 4,
1984
to October 2, 2009, respectively. The bottom left and right graphs in
Figure ~\ref{fig:FTSESP}
show the analogous sample extremograms applied to 4,848
daily log-returns of the DAX Index from November 13, 1990
to October 2, 2009 and
to 6,333 daily log-returns of the Nikkei 225 Index
from August 23, 1984 to October 2, 2009, respectively.\footnote{As noted in the literature, the lower tails of returns tend to be heavier than the upper tails.  Similar plots (not shown here) of the extremogram for the upper tails also reveal extremal dependence, but to a lesser extent than seen in the lower tails.}
The daily
log-returns were calculated from the daily closing prices.
Notice that the extremograms for all four indices decay rather slowly to zero,
with S\&P the slowest.  Again the solid horizontal lines are 98\% permutation produced bounds. Interestingly, the
first lag in the sample extremogram for three of the indices (FTSE, DAX and Nikkei)
is smaller than the second lag. Among the four indices, the Nikkei displays the least amount of extremal dependence as measured by the extremogram.

\begin{figure}[ht]
\begin{center}
\centerline{\includegraphics[height=8cm,width=16cm]{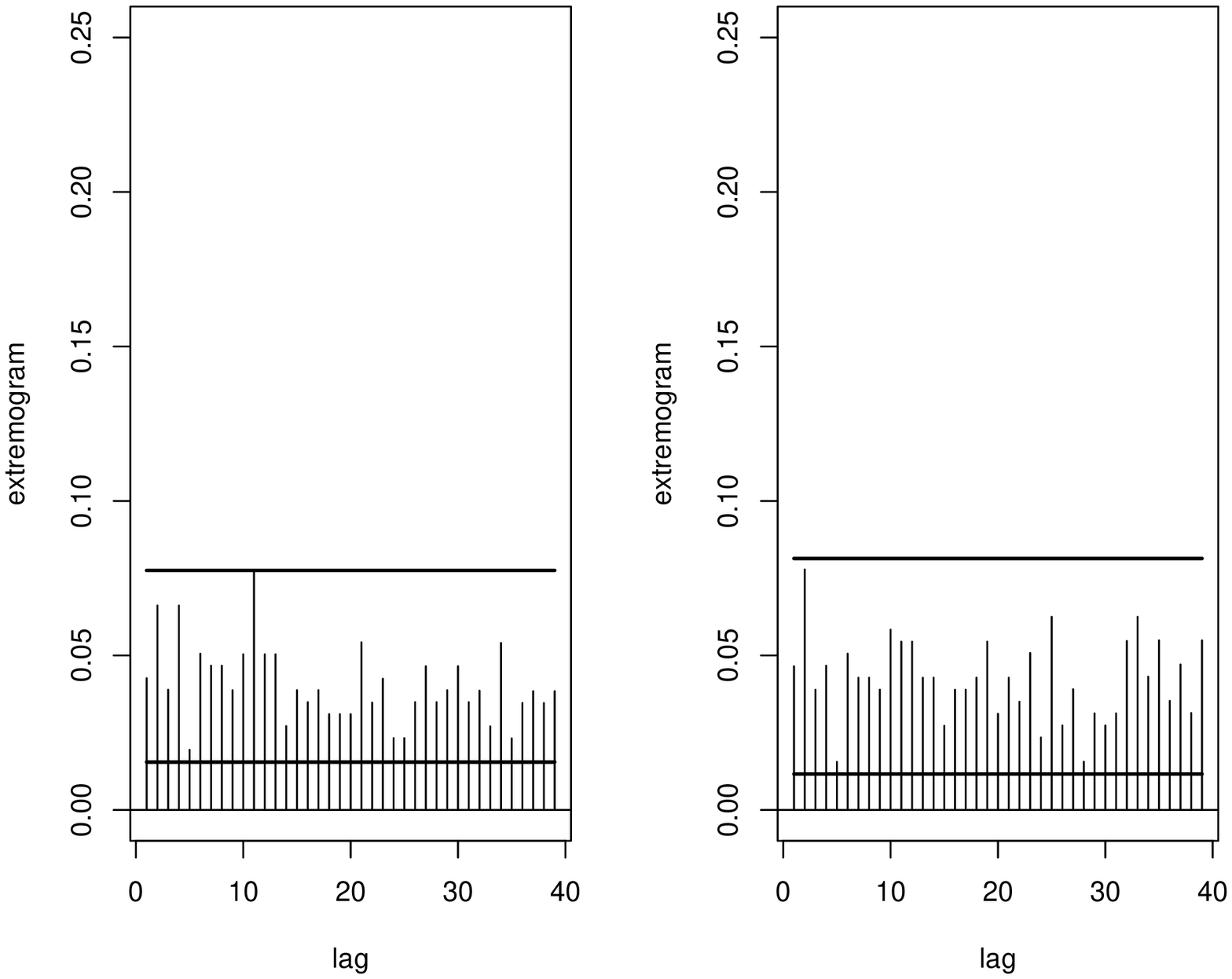}}
\end{center}
\bfi{\small The sample extremograms for the
filtered FTSE {\rm (left)} and the filtered S\&P {\rm (right)}. The bold lines represent sample extremograms
based on $99$ random permutations of their respective filtered series.}
\label{fig:FTSESPres}
\efi
\end{figure}

The top graphs in
Figure~\ref{fig:FTSESP} indicate extremal dependence in the lower tail over a period of
40 days. Typically, financial returns are modeled via a multiplication model, given by
$X_t=\sigma_tZ_t$,
where $(Z_t)$ is an iid sequence of mean 0 variance 1 random variables
and for each $t$, the volatility $\sigma_t$ is independent of $Z_t$.
It is often assumed that $(Z_t)$ is heavy-tailed as well.  Most financial time series models
such as GARCH and SV, have this
form.  With such a model, an extreme value of the process occurs at
time $t$ if $\sigma_t$ is large or if there is a large {\it
  shock} in the noise (i.e., $Z_t$) at time $t$. After estimating the
volatility process $(\sigma_t)$, the estimated devolatilized process
is defined by $\hat Z_t=X_t/\hat\sigma_t$.  If the multiplicative
model is a reasonable approximation to the return series, then the
devolatilized series should be free of extremal  dependence.  For each
of the four indices, FTSE, S\&P, DAX, and Nikkei, (denoted by $X_{t1},
X_{t2}, X_{t3}$ and $X_{t4}$, respectively) a GARCH(1,1) was used to
devolatilize each of the four component series.
Let $\hat Z_{ti}=X_{ti}/\hat \sigma_{ti}, i=1,\ldots,4,$ be the respective devolatilized series.
Figure~\ref{fig:FTSESPres}
shows the sample extremograms and bounds produced by the permutation procedure of the left tail
for the filtered series $\hat Z_{t1}$ and $\hat Z_{t2}$ corresponding to FTSE and S\&P.  Here we used $A = B = (-\infty,-1)$ and the negative of the .04 empirical quantile for $a_m$.  These plots confirm that the extremal dependence as measured by the extremogram has been removed.
Results for the filtered DAX and Nikkei are very similar.  Thus the extremal dependence in the series $(X_{ti})$ is due solely to the persistence in the volatility series.

\section{Further Extensions.}\label{sec:further}
\subsection{Cross-extremogram for bivariate time series.}\label{sec:bivext}
While the definition of the extremogram in \eqref{eq:ext} covers the case of multivariate time series, it is implicit that the index of regular variation is the same across the component series.  For example, consider two regularly varying time series $(X_t)$ and $(Y_t)$ with tail indices $\alpha_1$ and $\alpha_2$ with $\alpha_1<\alpha_2$.  Then the bivariate time series $((X_t,Y_t)')_{t\in\bbz}$ would be regularly varying with index $\alpha_1$ and
$$
\lim_{x\to\infty}P(x^{-1}Y_{t+h}\in  B \mid x^{-1}X_t\in  A)=0.
$$
In this case, the extremogram involving the $Y_t$ series would be zero
and there would be no extremal dependence between the two component series.  To avoid these rather uninteresting cases and obtain a more meaningful measure of extremal dependence, we transform the two series so that they have the same marginals.  In extreme value theory, the transformation to the unit Fr\'echet distribution is the most standard.  For the sake of argument in this discussion, assume that both $X_t$ and $Y_t$ are positive so that the focus of attention will be on extremal dependence in the upper tails.  The case of extremal dependence in the lower tails or upper and lower tails is similar.  Under the positivity constraint, if $F_1$ and $F_2$ denote the marginal distributions of $X_t$ and $Y_t$, respectively, then the two transformed series, $\tilde X_t=G_1(X_t)$ and $\tilde Y_t=G_2(Y_t)$, have unit Fr\'echet marginals ($F(x)=\exp\{-1/x\}$, $x>0$), where $G_i(z)=-1/\log(F_i(z))$.  Now assuming that the bivariate time series $((\tilde X_t,\tilde Y_t)')_{t\in\bbz}$ is regularly varying, we define the {\it cross-extremogram} by
\beao
\rho_{A,B}(h)=\lim_{x\to\infty}P(x^{-1}\tilde Y_{h}\in B\mid x^{-1}\tilde X_0 \in
A)\,,\quad
h\ge 0\,,
\eeao
where $A$ and $B$ are sets bounded away from 0.
\par
At first glance, this may seem unpleasant since transformations to unit Fr\'echets are required.  However if one restricts attention to sets $A$ and $B$ that are intervals bounded away from 0 or finite unions of such sets, then since  $n$ is the $(1-1/n)$-quantile of a Fr\'echet distribution, we have by the monotonicity of the transformation $G_i$, $\{n^{-1}\tilde X_h\in A\}=\{a_{n,X}^{-1}X_h\in A\}$ and $\{n^{-1}\tilde Y_h\in B\}=\{a_{n,Y}^{-1}Y_h\in B\}$, where $a_{n,X}$ and $a_{n,Y}$ are the respective $(1-1/n)$-quantiles of the distributions of $X_t$ and $Y_t$.  So as long as the sets $A$ and $B$ have this form, the cross-extremogram becomes
\beam\label{eq:crossext}
\rho_{A,B}(h)&=&\lim_{n\to\infty}P(a_{n,Y}^{-1}Y_h\in B\mid  a_{n,X}^{-1}X_0 \in A)\,.
\eeam
The point of this observation is that
{\em we do not need to actually find the transformations converting
  the data to
unit Fr\'echets, only the component-wise quantiles, $a_{n,X}$ and $a_{n,Y}$, need to be calculated.}
Clearly, this notion of extremogram extends to more than two time series.

\subsection{Sample cross-extremogram for bivariate time series.}\label{sec:bivext2}
As argued in Section \ref{sec:bivext}, the cross-extremogram for the
bivariate time series $((X_t,Y_t))_{t\in  \bbz}$ is
\beam \label{eq:bivpopext}
\rho_{A,B}(h)=\lim_{m\to\infty}\rho_{A,B:m}(h)=\lim_{x\to\infty}P(a_{m,Y}^{-1}Y_{h}\in B\mid a_{m,X}^{-1}X_0 \in A)\,,\quad
h\ge 0\,,
\eeam
where $a_{m,Y}$ and $a_{m,X}$ are the $(1-m/n)$-quantiles of the distributions of $|Y_t|$ and $|X_t|$, respectively, and
$A$ and $B$ are finite unions of intervals that are bounded away from 0. (In cases where we only explore the upper or lower tails, we then choose $a_{m,X}, a_{m,Y}$ to be either the $m/n$ or the $(1-m/n)$-quantiles of the respective distribution functions.)  For the sake of simplicity,
we use the same symbol $\rho_{A,B}(h)$ as before, abusing notation and refer to $\rho_{A,B:m}$ as the {\it pre-asymptotic cross-extremogram}.

Starting from \eqref{eq:bivpopext}, we define the
{\em sample cross-extremogram} for the time series $((X_t,Y_t))_{t\in\bbz}$ by
\beam\label{eq:bivsamext}
\wh\rho_{A,B}(h)=\frac{\sum_{t=1}^{n-h}I_{\{a_{m,Y}^{-1}Y_{t+h}\in B,a_{m,X}^{-1}X_t\in A\}}}{\sum_{t=1}^{n}I_{\{a_{m,X}^{-1}X_t\in A\}}}\,,
\eeam
where $a_{m,X}$ and $ a_{m,Y}$ are replaced by the respective empirical quantiles computed from $(X_t)_{t=1,\ldots,n}$ and $(Y_t)_{t=1,\ldots,n}$,
respectively.
\par
We are interested in the extremal serial dependence in the left tail
for the pairs of equity index log-returns
discussed in Section~\ref{sec:EQIndex}.
Therefore we calculate
the sample cross-extremograms
for these pairs with sets
$A = B = (-\infty,-1)$ and the negative $.04$ empirical quantiles of the returns for
$a_{m,X}$ and $a_{m,Y}$, respectively.
\par
\begin{figure}[ht]
\begin{center}
\centerline{\includegraphics[height=4cm,width=16cm]{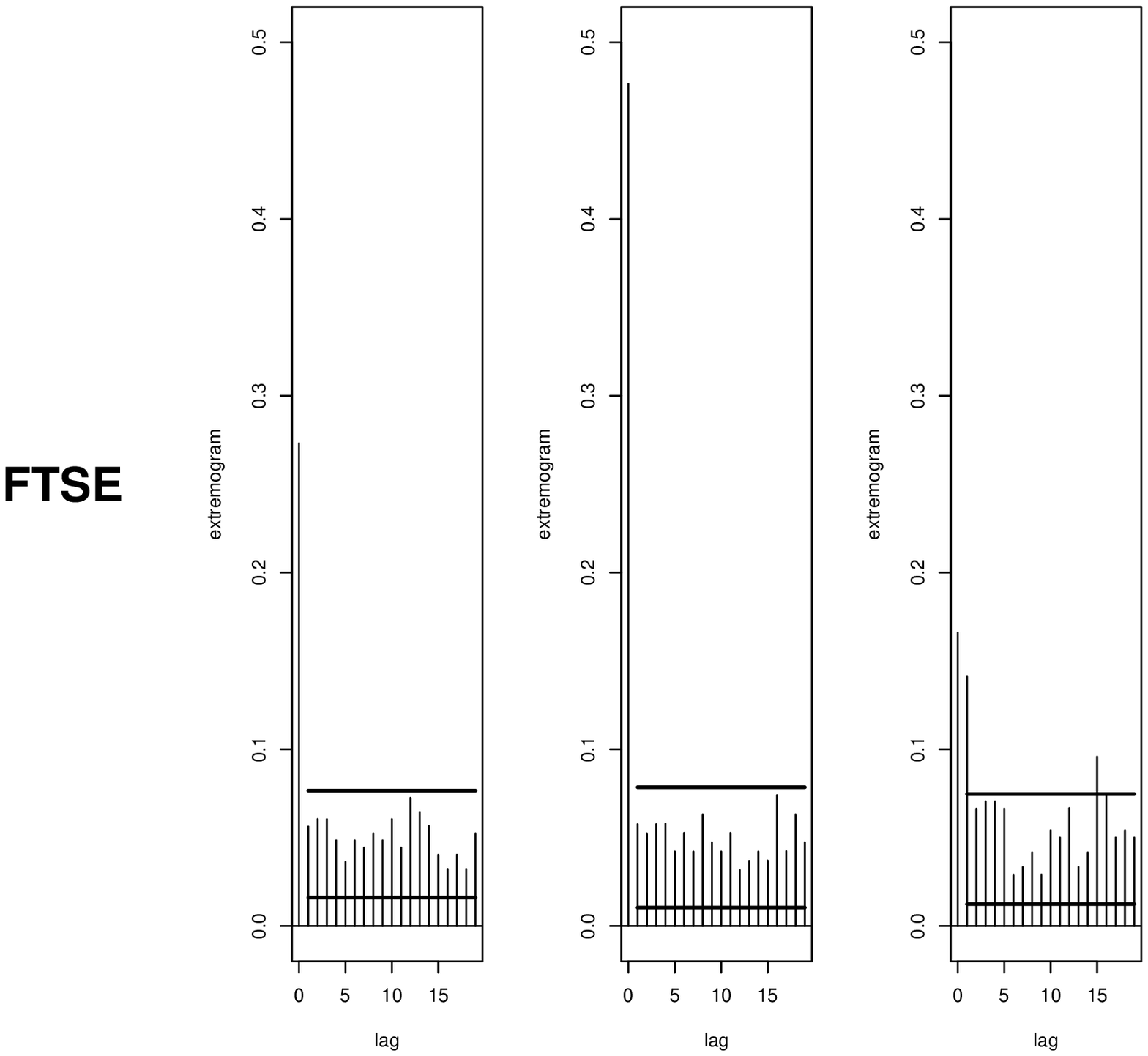}}
\end{center}
\begin{center}
\centerline{\includegraphics[height=4cm,width=16cm]{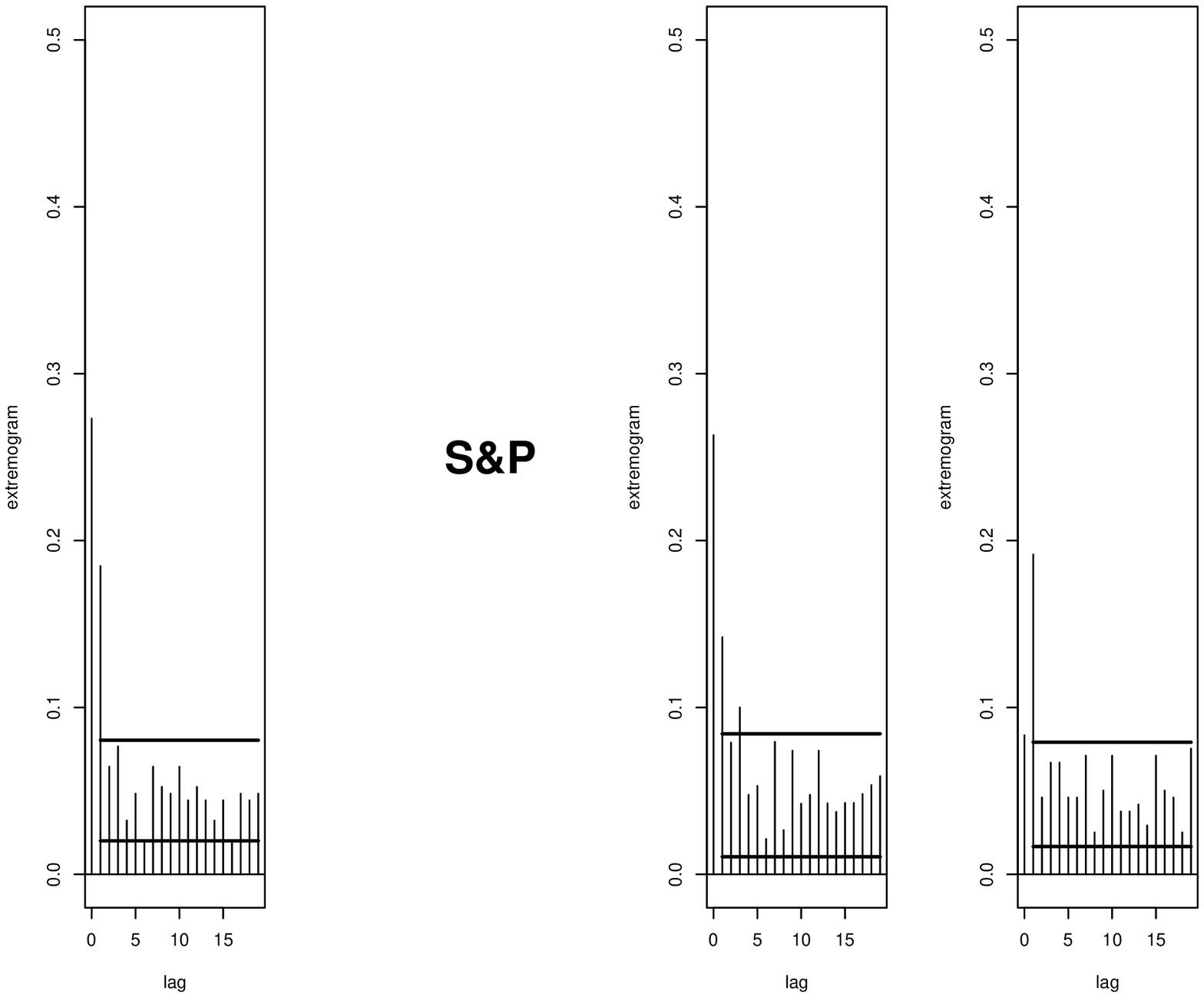}}
\end{center}
\begin{center}
\centerline{\includegraphics[height=4cm,width=16cm]{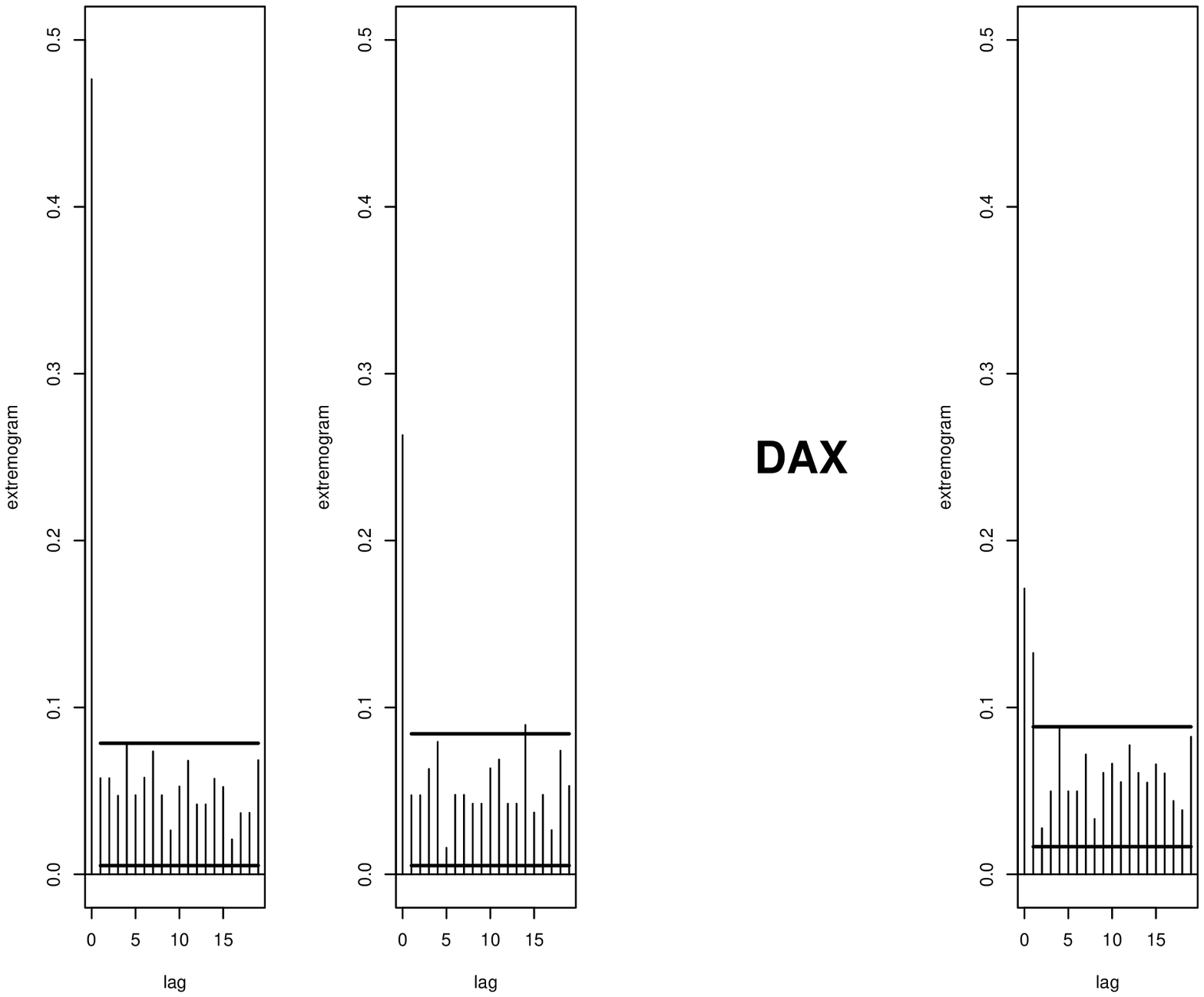}}
\end{center}
\begin{center}
\centerline{\includegraphics[height=4cm,width=16cm]{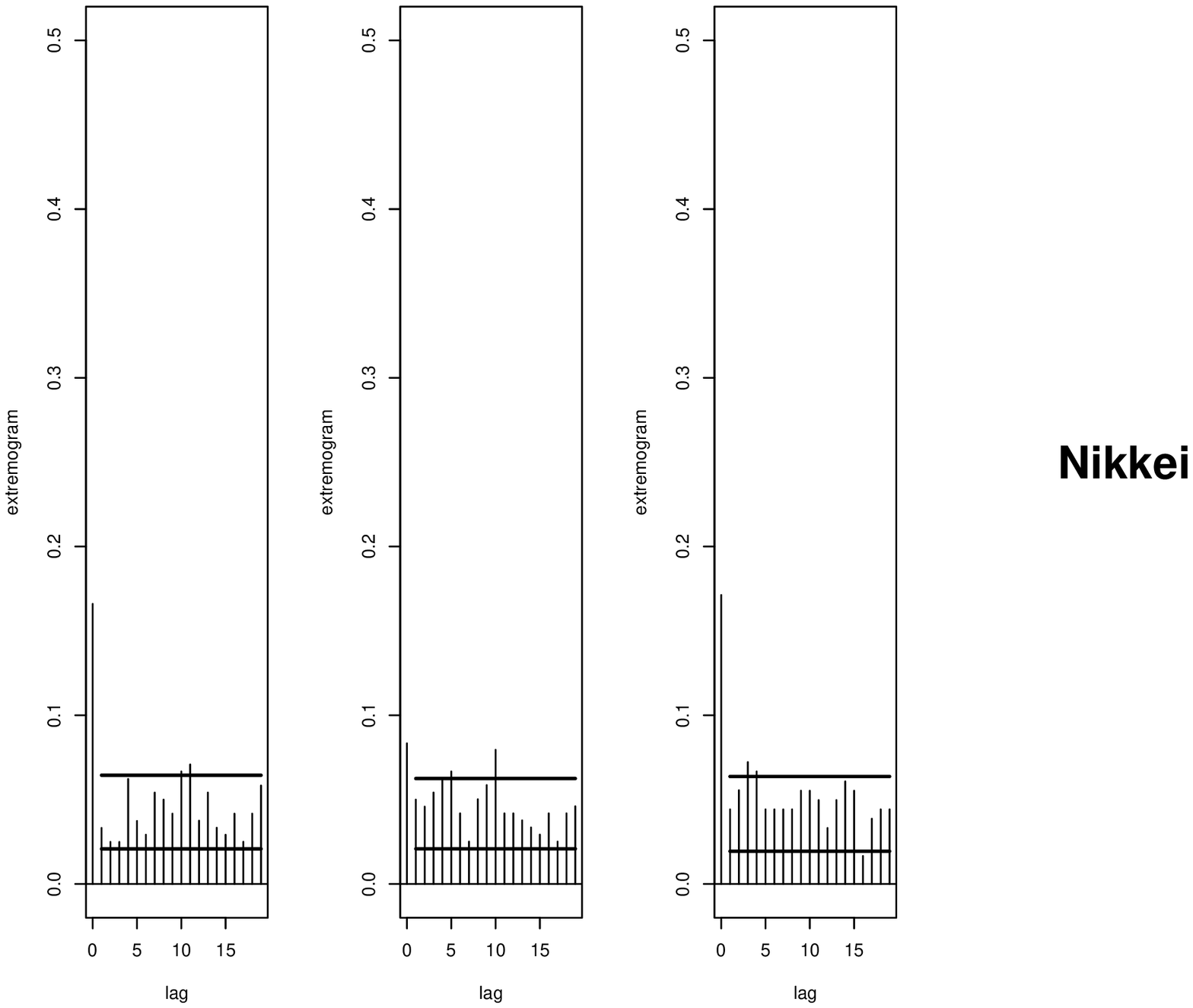}}
\end{center}
\bfi{\small The sample cross-extremograms for the filtered
FTSE, S\&P, DAX and Nikkei series. For the first row, $(X_t)$
is the filtered FTSE and $(Y_t)$ are the filtered S\&P, DAX and Nikkei
{(from left to right)}. For the second, third and fourth rows, the $X_t$'s
are the filtered S\&P, DAX and Nikkei series, respectively.}\label{fig:bivst}\efi
\end{figure}

In calculating the cross-extremograms between the daily indices
$(X_{ti})$ and $(X_{tj})$, we only consider days $t$ for which we have
observations on both indices.  Since FTSE, S\&P, DAX, and Nikkei are
indices from four different countries, there is not a perfect overlap
on when the corresponding markets are open.  The cross-extremograms
between $(X_{ti})$ and $(X_{tj})$ exhibited a similar pattern of slow
decay as seen in the univariate sample extremograms. It was thought
that the   strong cross-extremal dependence could be an artifact due
to the persistence and synchronicity in the marginal volatilities.
This phenomenon is similar in spirit to the well-known issue for
the cross-correlation function of linear bivariate time series.  In this case, unless one or all of the component time series have been whitened, the cross-correlation may appear to be significant (see Chapter 11 in Brockwell and Davis \cite{brockwell:davis:1991}).  To explore this phenomenon, we computed the cross-extremogram for the devolatilized components.  Figure~\ref{fig:bivst} shows the sample cross-extremogram  between the estimated residuals, $\hat Z_{ti}$ and $\hat Z_{tj}$ for $i\ne j$ corresponding to the four indices.  For example, in the first row
of graphs, $(X_t)$ is the filtered FTSE ($\hat Z_{t1}$) and
$(Y_t)$ are the filtered S\&P ($\hat Z_{t2}$), DAX ($\hat Z_{t3}$) and Nikkei ($\hat Z_{t4}$), respectively.  Each graph contains permutation generated confidence bands.

Interestingly, there are signs of various types of cross-extremal dependence in the filtered series.  The spike at lag zero (except between the Nikkei and S\&P)
indicates the extremal dependence in the shocks $\hat Z_{ti}$ and $\hat Z_{tj}$ for $i,j=1,2,3$.
This is not surprising since we would expect dependence (extremal or otherwise) between the devolatized series obtained from univariate GARCH(1,1) fits to each of the marginal series.
In the second row,
there is evidence of significant extremal dependence at
lag one for each sample cross-extremogram: given the
S\&P has an extreme left tail shock at time $t$
there will be a corresponding large left tail shock in the FTSE, the DAX and the Nikkei
at time $t+1$.  Given the dominance of the US stock market, one might expect a carry-over effect of the shocks on the other exchanges on the next day.  Since only marginal GARCH models were fitted to the data, it may not seem all that surprising that the filtered series exhibit serial dependence.  We should note, however, that the dependence in the shocks does not appear to last beyond one time lag.

We also computed the extremogram between $(X_{t1})$ and $(\hat Z_{ti})$ for $i=2,3,4$.  These plots (not shown) were virtually identical to those displayed in Figure~\ref{fig:bivst} which suggests that only one of the components in the cross-extremogram needs to be devolatized.

\par
While this analysis
was carried out on the left tail, the right tail (not included)
shows very similar patterns. However, the degree of dependence is
different: the left tail extremal dependence probability is greater
than in the right tail.

\subsection{Cross-extremogram for trivariate time series.}\label{sec:Triext}
For a stationary trivariate \regvary\ time series
$((X_t,Y_t,Z_t))_{t\in\bbz}$, many different variations of the
cross-extremogram can be defined depending on the context. We focus on the extremograms
\beam \label{eq:tri2ext}
\rho_{1}(h)&=&\lim_{x\to\infty}P(\{Y_{h}>x\}\cup
\{Z_{h}>x\}\mid X_0>x)\,,\quad h\ge 0\,,\\
\label{eq:tri3ext}
\rho_{2}(h)&=&\lim_{x\to\infty}P(Z_{h}>x\mid \{Y_{0}>x
\}\cup\{ X_0>x\})\,,\quad h\ge 0\,.
\eeam
Similar to the discussion in Section~\ref{sec:bivext}, one needs to
replace the thresholds $x$ by the quantiles of the marginal \ds s
if these are not identical. The
sample cross-extremograms corresponding to
\eqref{eq:tri2ext} and \eqref{eq:tri3ext}, respectively, are then
defined as
\beao
\wh\rho_{1}(h)=\dfrac{\sum_{t=1}^{n-h} I_{\{
X_t>a_{m,1}\;\mbox{and}\;
(Y_{t+h}>a_{m,2}\;\mbox{or}\; Z_{t+h}>a_{m,3})\}}}{\sum_{t=1}^{n}I_{\{X_t>
a_{m,1}\}}}\,,\quad h\ge 0\,,\\
\wh\rho_{2}(h)=\dfrac{\sum_{t=1}^{n-h} I_{\{
(X_t>a_{m,1}\;\mbox{or}\; Y_{t}>a_{m,2})\;\mbox{and}\; Z_{t+h}>a_{m,3})\}}}{\sum_{t=1}^{n}I_{\{X_t>a_{m,1}\;\mbox{or}\; Y_{t}>a_{m,2}
\}}}\,,\quad h\ge 0\,,
\eeao
$a_{m,i}$, $i=1,2,3,$ are chosen as the
corresponding empirical quantiles of the $X_t$'s,
$Y_t$'s and $Z_t$'s, respectively.
\par
\begin{figure}[ht]
\begin{center}
\centerline{\includegraphics[height=8cm,width=16cm]{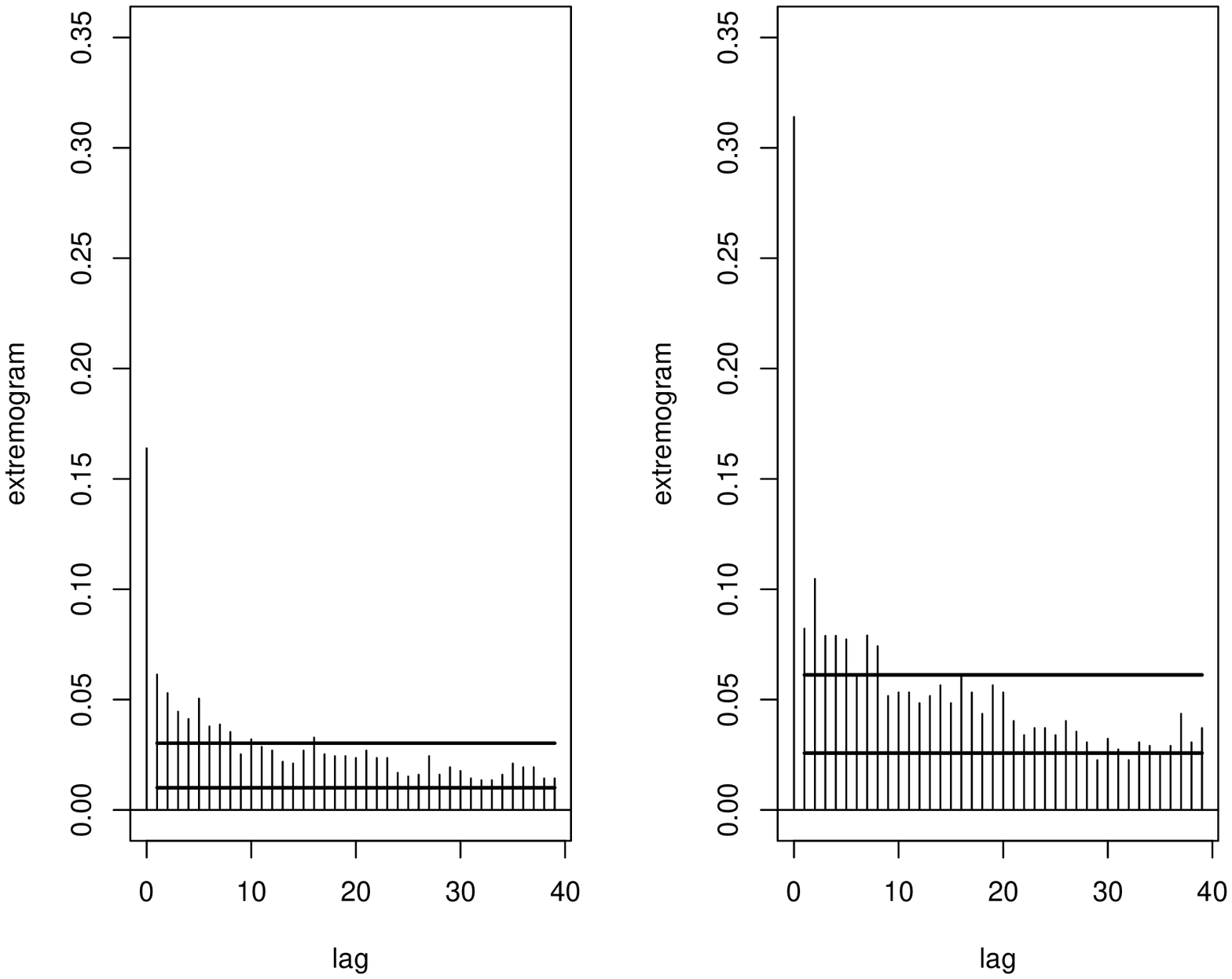}}
\end{center}
\bfi{\small The sample cross-extremograms for
\eqref{eq:tri2ext} {\rm (left)}
and \eqref{eq:tri3ext} {\rm (right)}
for the $5$-minute log-returns $X_t,Y_t,Z_t$ of BAC, CBK and MSFT, respectively.}
\label{fig:trist}\efi
\end{figure}

Figure ~\ref{fig:trist} shows the sample cross-extremograms
corresponding to \eqref{eq:tri2ext} and \eqref{eq:tri3ext}.
In both graphs, $(X_t)$, $(Y_t)$ and $(Z_t)$ represent the 5-minute
log-returns of Bank of America (BAC), Citibank (CBK) and Microsoft
(MSFT) from December 1, 2004 to July 26, 2006, respectively.
Here, the absolute values of the $.96$ empirical
quantiles of the  negative returns for the series
$(X_t)$, $(Y_t)$ and $(Z_t)$ are used for
$a_{m,1}$, $a_{m,2}$ and $a_{m,3}$, respectively.
In the left graph of Figure ~\ref{fig:trist}, one can interpret the
spike at lag zero as the probability
of obtaining an extreme return in either CBK or MSFT now given that
there is an extreme return in BAC now.
The spike at lag one has the interpretation as the probability of
obtaining an extreme return in either CBK or MSFT now
given that there was an extreme return in BAC 5 minutes ago.
The interpretation of the spikes as lags 0 and 1 in the right graph is
analogous. In both plots the decay is fast. If one wanted to
utilize the information about the extremal serial dependence between
the series
one would have to act fast.

\subsection{The extremogram of return times between rare events}\label{sec:Geman}\setcounter{equation}{0}
In their presentation \cite{geman:chang:2009},
Geman and Chang consider the waiting or return
times between rare (or
extreme) events for financial time series.
They define a rare event by a large excursion
relative to observed returns: a return $X_t$ is {\em rare} (or {\em
  extreme})
if $X_t \le \xi_{p}$ or $X_t \ge \xi_{1-p}$,
where $\xi_{q}$ is the $q$-quantile of the distribution of
returns. Typical choices for $p$ are 0.1 and 0.05.
Denoting occurrences of rare events by the binary sequence
\beam
W_j =\left\{\barr{ll}
  1, & \mbox{if $X_t$ is extreme,}\\
  0, & \mbox{otherwise,}
\earr\right.
\eeam
Geman and Chang study return times $T_j$, $j=1,2,\ldots$,
between successive 1's
of the $W_j$ sequence.
If the return times
were truly iid, the successive
waiting times between 1's should be iid
geometric. Using the histogram of waiting times,
the geometric assumption can be examined.
In order to perform inference and
in particular hypothesis testing
on the extremal clustering of returns, Geman and Chang
\cite{geman:chang:2009} calculate
the observed entropy of the
excursion waiting times
and compare it to the entropies of
random permutations of the excursion
waiting times.
If the observed entropy behaves similarly
to the random permutations
then one may conclude that the returns do not exhibit extremal clustering.
\par
We now introduce an analog of the extremogram for the return times
between
rare events in a strictly stationary \regvary\ $\bbr^d$-valued \seq\ $(X_t)$. Denoting the rare
event by $A\subset \ov \bbr_0^d$, the corresponding return times extremogram is given by
\beam\label{eq:rhotime}
\rho_{A}(h)=\lim_{\xto} P(X_1\not\in x A,\ldots,X_{h-1}\not\in
xA,X_h\in x A\mid
X_0\in x A)\,,\quad h\ge 1\,.
\eeam
Using the \regvar\ of the \seq\ $(X_t)$ (see Section~\ref{sec:ext})
and assuming that $A$ and $A\times \ov \bbr_0^{d(h-1)}\times A$ are
continuity sets \wrt\ $\mu$ and $\mu_{h+1}$, $h\ge 1$, and $A$ is bounded away
from zero, we can calculate the return times extremogram
\beao
\rho_{A}(h)=\dfrac{\mu_{h+1}(A\times \bbr^{h-1}\times
  A)}{\mu(A)}\,,\quad h\ge 0\,.
\eeao
The return times sample extremogram is then defined as
\beam \label{eq:GCext}
\wh\rho_{A}(h)=\dfrac{\sum_{t=1}^{n-h}I_{\{a_m^{-1}X_{t+h}\in
    A,a_m^{-1}X_{t+h-1}\not \in
A,\ldots, a_m^{-1}X_{t+1}\not \in A,a_m^{-1}X_t\in A\}}}
{\sum_{t=1}^{n}I_{\{a_m^{-1}X_t\in  A\}}},\quad h=1,2,\ldots,n-1.
\eeam
An \asy\ theory for the return times sample extremogram and its
bootstrap version is given at the end of Section~\ref{subsec:consist}.
This theory shows that the stationary bootstrap is \asy ally correct
for this sample extremogram.
\par
\begin{figure}[ht]
\begin{center}
\centerline{\includegraphics[height=8cm,width=16cm]{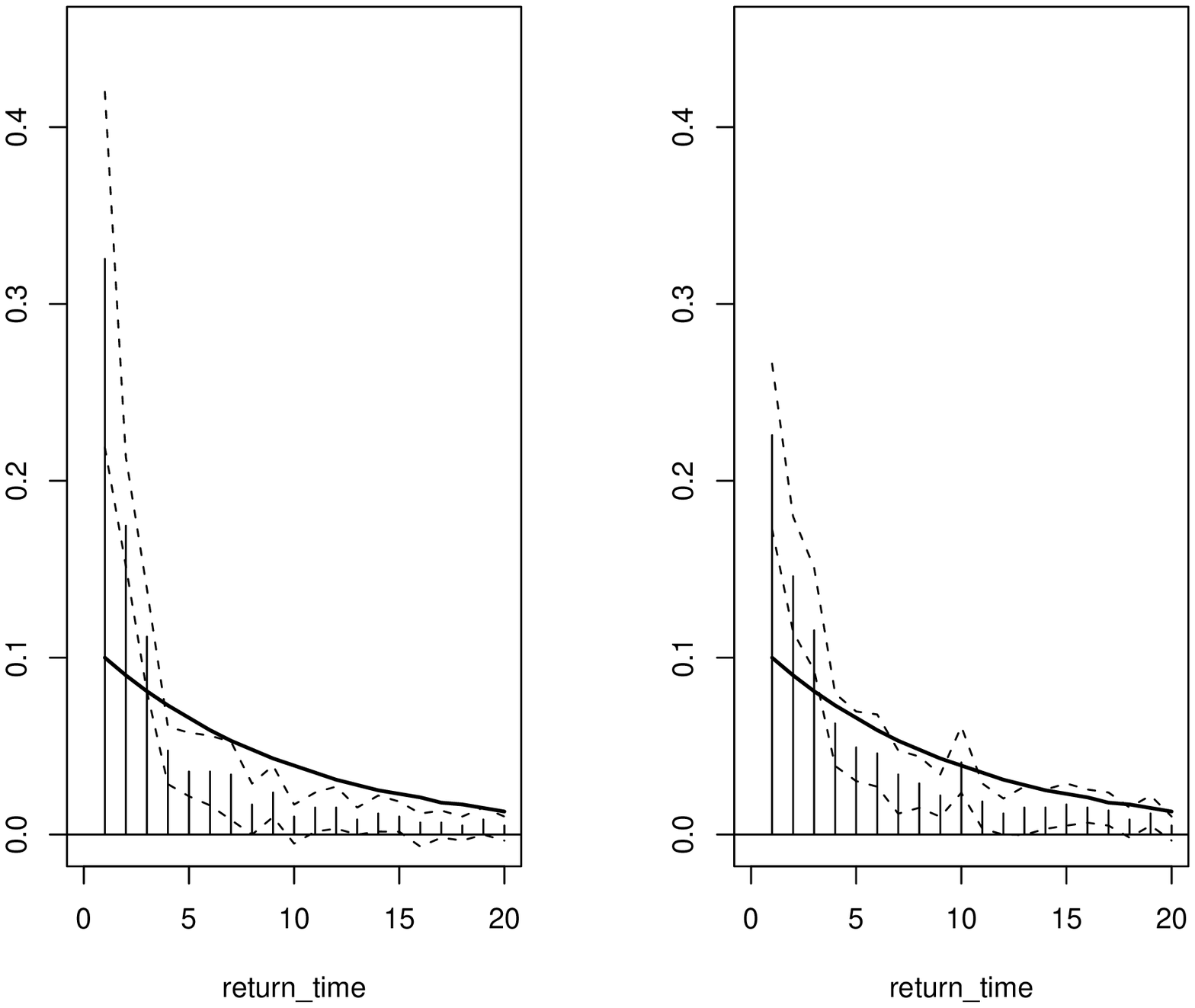}}
\end{center}
\begin{center}
\centerline{\includegraphics[height=8cm,width=16cm]{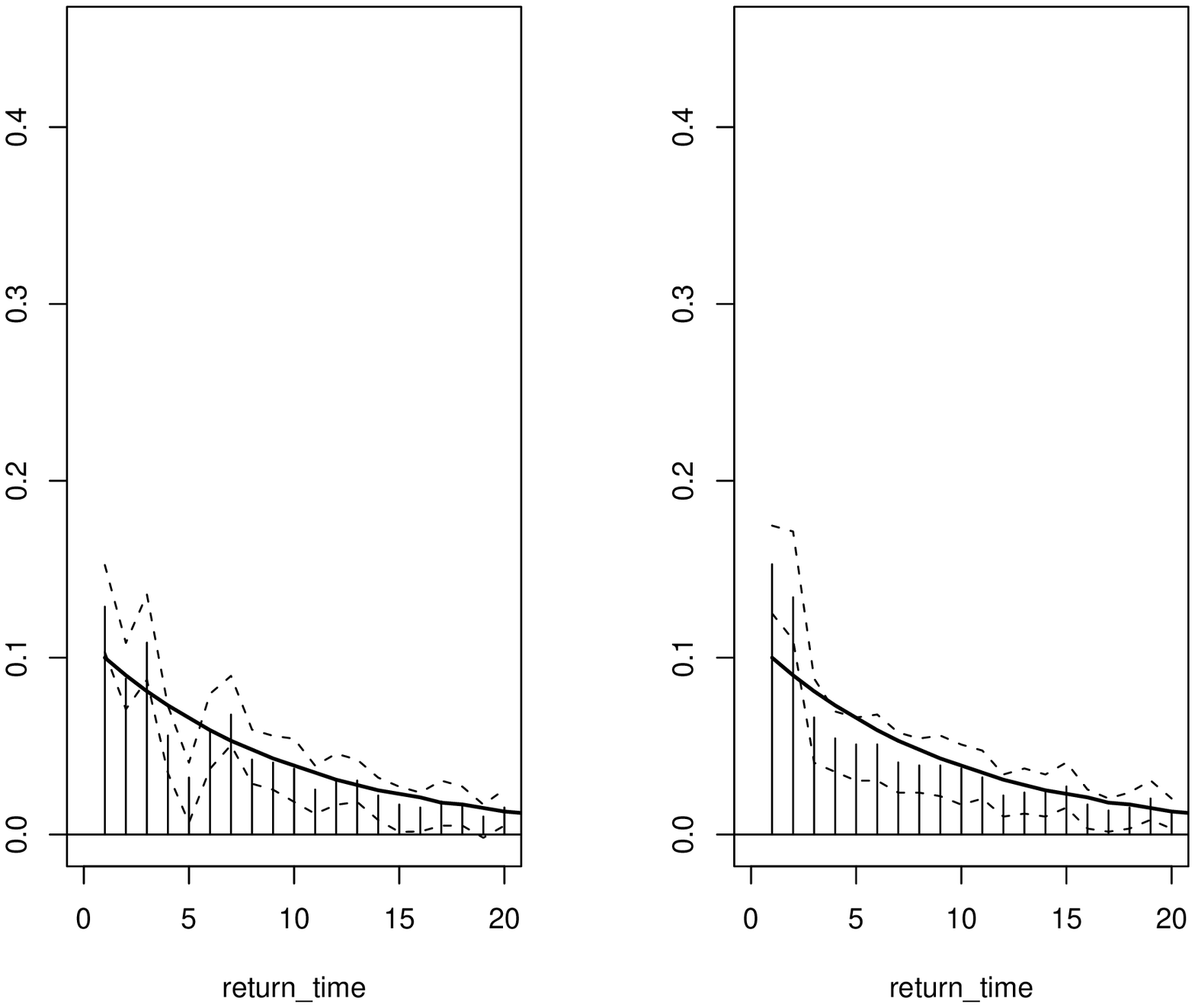}}
\end{center}
\bfi{\small
The histograms (solid vertical lines) for the return times of extreme events
for the daily log-returns of BAC using bootstrapped confidence intervals (dashed lines), geometric probability mass function (light solid)  for $A=\bbr\backslash
[\xi_{0.05},\xi_{0.95}]$
{\rm (left)} and
$A=(-\infty,\xi_{0.1})$ {\rm (right)}.}
\label{fig:geman}\efi
\end{figure}

We will now illustrate some examples of the return times of extreme events.
The graphs in Figure \ref{fig:geman} show the
histograms for the return times of extreme events for the daily
log-returns of Bank of America (BAC) for different choices of the
rare events $A$: in the left graphs $A=\bbr\backslash [\xi_{0.05},\xi_{0.95}]$
and in the right graphs $A=(-\infty,\xi_{0.1})$.
The top row contains the histogram (solid vertical lines) of the log-returns, whereas the
bottom row shows the corresponding histograms for the filtered \ts\
after fitting a GARCH(1,1) model to the data; cf. Section~\ref{sec:EQIndex}.  Since the heights of the histogram correspond exactly to the return times sample extremogram, we can apply the bootstrap procedures of Section~\ref{sec:bootext}.  The dashed lines that overlay the graphs in Figure \ref{fig:geman} represent the .975 (upper) and .025 (lower) confidence bands computed from the bootstrap approximation to the sampling distribution of the sample extremogram.  The solid curve is the  geometric probability mass function with success \pro y $p=0.1$.  As seen in these plots, the geometric probability mass function falls outside the confidence bands at nearly every value.  This complements earlier findings of the presence of serial extremal dependence in the original daily returns.

\section{Appendix: Proofs}\label{sec:proofs}\setcounter{equation}{0}

\subsection{Proof of Theorem~\ref{thm:1}}
Notice that
$E^\ast (\wh P_m^\ast)= \wh P_m$. Now it follows from Theorem 3.1 in
Davis and Mikosch \cite{davis:mikosch:2009} that
\beam\label{eq:11}
\wh P_m\stp
\mu(C)\,.
\eeam
This proves \eqref{eq:1}.
\par
Next we prove \eqref{eq:2a}, i.e., we
study the \asy\ behavior of $m\,s_n^2$. Write $\wt
I_t=I_t-p_0$ and
\beao
\wt C_n(h)&=&n^{-1} \sum_{i=1}^n \wt I_i\wt I_{i+h}\,,\quad
h=0,\ldots,n\,,\\[2mm]
\wt \gamma_n(h)&=&n^{-1} \sum_{i=1}^{n-h} \wt I_i\wt I_{i+h}\,,\quad
h=0,\ldots,n-1\,,
\eeao
where the $I_j$'s in the summands are again defined circularly, i.e.,
$I_j=I_{j\,{\rm mod}\, n}$.
Since
\beao
C_n(h)=n^{-1} \sum_{i=1}^n I_i\,I_{i+h}-(\ov I_n)^2=
\wt C_n(h) +p_0^2-(\ov I_n)^2
\eeao
and from the central limit theorem in (4) of Theorem~\ref{thm:1} and \eqref{eq:11},
\beao
(n/m)^{1/2}\,m((\ov I_n)^2-p_0^2)=
 \big[(n/m)^{1/2} \,(\wh P_m-m\,p_0)\big]\,\big[
m^{-1}(\wh P_m +m\,p_0)\big]=O_P(m^{-1})\,,
\eeao
we have
\beam
\lefteqn{m\,s_n^2-m\,\left(\wt C_n(0)+ 2\sum_{h=1}^{n-1}
(1-h/n)\,(1-p)^h\,\wt C_n(h)\right)}\nonumber\\[2mm]&=&m((\ov I_n)^2-p_0^2)
\,\left(1+ 2\sum_{h=1}^{n-1}
(1-h/n)\,(1-p)^h\right)\nonumber\\[2mm]
&=&O_P((mn)^{-1/2}p^{-1})=o_P(1)\,.\label{eq:aa}
\eeam
In the last step we used assumption \eqref{eq:9}.
It follows from \eqref{eq:5} and Lemma~5.2 in Davis and Mikosch
\cite{davis:mikosch:2009} that
\beam\label{eq:bb}
m\,\wt C_n(0)\stp \mu(C)\quad \mbox{and}\quad m\,\wt C_n(h)\stp \tau_h(C)\,,\quad
h\ge 1\,.
\eeam
We also have by assumption \eqref{eq:6},
\beam\label{eq:cc}
\lim_{k\to\infty}\limsup_{\nto}E\left|m\sum_{h=k}^{r_n}(1-h/n)\,(1-p)^h \wt C_n(h)\right|
&\le& \lim_{k\to\infty}\limsup_{\nto}m\,\sum_{h=k}^{r_n}p_{0h}=0\,.
\eeam
Combining \eqref{eq:aa}--\eqref{eq:cc} and recalling the definition of
$\sigma^2(C)$ from \eqref{eq:2},
it suffices for \eqref{eq:2a} to show that for every $\delta>0$,
\beao
\lim_{k\to\infty}\limsup_{\nto}P\left(m\,\left|\sum_{h=r_n}^{n-1}
(1-h/n)\,(1-p)^h\,\wt C_n(h)\right|>\delta\right)=0\,.
\eeao
We have
\beao
m\,\sum_{h=r_n}^{n-1}
(1-h/n)\,(1-p)^h\,\wt C_n(h)=\sum_{h=r_n}^{n-r_n}b_n(h)
\,\wt \gamma_n(h)+o_P(1)=T_n+o_p(1)\,,
\eeao
where
\beao
b_n(h)=m\,\left[(1-h/n)\,(1-p)^h+
(h/n)\,(1-p)^{n-h}\right]\,.
\eeao
Then
\beam
\var(T_n)&=&\sum_{h_1=r_n}^{n-r_n}\sum_{h_2=r_n}^{n-r_n}
b_n(h_1)b_n(h_2)\,\cov(\wt \gamma_n(h_1),\wt \gamma_n(h_2))\nonumber\\[2mm]
&\le &\sum_{h_1=r_n}^{n-1}\sum_{h_2=r_n}^{n-1}
b_n(h_1)b_n(h_2) \max_{r_n\le h_1,h_2<n-1}| \cov(\wt \gamma_n(h_1),\wt
\gamma_n(h_2))|\nonumber\\[2mm]
&\le& c\,p^{-2}m^2\max_{r_n\le h_1,h_2<n-1}| \cov(\wt \gamma_n(h_1),\wt
\gamma_n(h_2))|\,.\label{eq:8}
\eeam
Here and in what follows, $c$ denotes any positive constants whose
value is not of interest.
We have for fixed $k$,
\beao
m^2|\cov(\wt \gamma_n(h_1),\wt
\gamma_n(h_2))|&=&(m/n)^2\left|\sum_{t=1}^{n-h_1}\sum_{s=1}^{n-h_2}\cov(I_tI_{t+h_1},I_sI_{s+h_2})\right|\\[2mm]
&\le &c\,m^2/n\sum_{r=1}^{n}|\cov(I_0I_{h_1},I_rI_{r+h_2})|\\[2mm]&=&
c\,(m/n)\,m\left(\sum_{|r-h_1|\le k}+\sum_{k<|r-h_1|\le
    r_n}+\sum_{|r-h_1|>r_n}\right)|\cov(I_0I_{h_1},I_rI_{r+h_2})|\\[2mm]
&=&c\,(m/n)[J_1+J_2+J_3]\,.
\eeao
Using the sequential definition of the \regvar\ of $(X_t)$ (see \eqref{eq:regvar3}),
we have for fixed $k$
\beao
\limsup_{\nto} J_1\le \limsup_{\nto}\sum_{|r-h_1|\le k} m\,[p_{0,|r-h_1|}+p_0^2]\le
\sum_{h\le k} \tau_h(C)\le \sigma^2(C)<\infty\,.
\eeao
Next we use condition \eqref{eq:6}:
\beao
\lim_{k\to\infty}\limsup_{\nto}J_2\le
\lim_{k\to\infty}\limsup_{\nto} m\, \sum_{k<|r-h_1|\le
    r_n} p_{0,|r-h_1|} +c\,\lim_{\nto}m\,r_n\,p_0^2=0\,.
\eeao
Finally, the mixing condition \eqref{eq:7} yields
\beao
\limsup_{\nto} J_3\le c\,\limsup_{\nto}m\,\sum_{|r-h_1|>r_n} \alpha_{|r-h_1|}=0\,.
\eeao
Thus we proved that
\beao
m^2|\cov(\wt \gamma_n(h_1),\wt
\gamma_n(h_2))|\le c \,(m/n)
\eeao
uniformly for $h_1,h_2\ge r_n$ and large $n$. We conclude from \eqref{eq:8}
and assumption \eqref{eq:9} that
\beao
\var(T_n)\le c\, m/(np^2)\to 0\,.
\eeao
Using \eqref{eq:7}, it also follows that $ET_n\to 0$. Thus we proved
that $E(T_n^2)\to 0$. Combining the bounds above, we conclude that
\eqref{eq:2a} is satisfied.
\par
Next we prove the \clt\ \eqref{eq:3}. Since both sums $\wh P_m^\ast$
and $\wh P_m$ contain the same number of summands and we
consider the difference
$(n/m)^{1/2}(\wh P^\ast_n-\wh P_m)$ we will assume in what
follows that all summands $(m/n)I_t$ in $\wh P_m^\ast$
and $\wh P_m$ are replaced by their centered versions $(m/n)\wt
I_t=(m/n)(I_t-p_0)$. We write $\wt P_m^\ast$ and $\wt P_m$ for the
corresponding centered versions.
\par
Write
\beam
S_{K_i,L_i}&=& \wt I_{K_i}+\cdots +\wt I_{K_i+L_i-1}\,,\quad i=1,2,\ldots,\nonumber\\
\label{eq:snm}
S_{nN}&=&S_{K_1,L_1}+\cdots + S_{K_N,L_N}\,.
\eeam
\ble\label{lem:1}
Under the conditions of Theorem~\ref{thm:1},
\beam\label{eq:14}
P^\ast\big((n/m)^{1/2}|\wt P_m^\ast-
(m/n) S_{nN}|>\delta\big)\to 0\,,\quad \delta>0\,.
\eeam
\ele
\begin{proof}
By the argument in Politis and Romano \cite{politis:romano:1994} on
p. 1312,
using the memoryless property of the geometric \ds ,
$
(m/n) S_{nN}-\wt P_m^\ast$ has the same \ds\ as $(m/n)S_{K_1,L_1}$
\wrt\ $P^\ast$.
Hence it suffices for \eqref{eq:14} to show that
$(m/n)E^\ast(|S_{K_1,L_1}|^2)\stp 0$. An application of Markov's
inequality shows that the latter condition is
satisfied if
\beam\label{eq:15}
(m/n) E( |S_{K_1,L_1}|^2)\to 0\,.
\eeam
We have by stationarity that
\beao
(m/n) E( |S_{K_1,L_1}|^2)&=& (m/n)\, E(|\wt I_1+\cdots
+\wt I_{L}|^2)\\[2mm]
&=& (m/n) \sum_{l=1}^{\infty} \var(I_{1}+\cdots +I_{l})\,(1-p)^{l-1}\,p\,.
\eeao
We have (sums over empty index sets being zero) for every fixed $k,l
\ge 1$,
\beao
m\,\var(I_{1}+\cdots +I_{l})&=&l\,\Big[
  m\,\var(I_1)+2 m\,\sum_{h=1}^{l-1}
  (1-h/l)\,\cov(I_0,I_h)\Big]\\[2mm]
&=&l\,\Big[
  m\,\var(I_1)+2 m\,\Big(\sum_{1\le h\le k}+\sum_{k<h\le
      r_n}+\sum_{r_n<h\le l-1}\Big)
  (1-h/l)\,\cov(I_0,I_h)\Big]\,.
\eeao
The \rhs\ is $O(l)$ uniformly for $l$ by virtue of \regvar\ of $(X_t)$ and in view of
the mixing condition (M). Since $np\to\infty$ we have
\beao
(m/n) E( |S_{K_1,L_1}|^2) \le c \,n^{-1}\sum_{l=1}^\infty l\,
(1-p)^{l-1}\,p
\le c\, (np)^{-1}\to 0\,.
\eeao
This proves \eqref{eq:15} and finishes the proof of the lemma.
\end{proof}
\noindent
In view of \eqref{eq:14} it suffices for \eqref{eq:3} to show that
$(n/m)^{1/2} [(m/n)S_{nN}-\wt P_m]\std N(0,\sigma^2(C))$
conditional on $(X_i)$. Recall that $E^\ast(S_{K_1,L_1})=p^{-1}(\ov
I_n-p_0)$. Then
\beam\label{eq:hilf}
&&(n/m)^{1/2} \left[(m/n)S_{nN}-\wt P_m\right]-
(m/n)^{1/2} \sum_{i=1}^N (S_{K_i,L_i}- p^{-1}(\ov I_n-p_0))\\[2mm]&&=
(n\,p)^{-1/2}\dfrac{N-n\,p}{\sqrt{n\,p}}\,\big[(n/m)^{1/2}\wt P_m
\big]\nonumber
\eeam
The quantities $((N-np)/\sqrt{np})$ are \asy ally normal as
$np\to\infty$ and have variances which are bounded for all $n$.
Therefore these quantities are stochastically bounded
in $P^\ast$-probability. Moreover, the quantities
$(n/m)^{1/2}\wt P_m$ have bounded variances in $P$-\pro y, and
therefore
\beao
(np)^{-1/2}[(n/m)^{1/2}\wt P_m]\stp 0\,.
\eeao
These arguments applied to \eqref{eq:hilf} yield for $\delta>0$,
\beao
P^\ast\Big(\Big|(n/m)^{1/2} \big[(m/n)S_{nN}-\wt P_m\big]-
(m/n)^{1/2} \sum_{i=1}^N (S_{K_i,L_i}- p^{-1}(\ov
I_n-p_0))\Big|>\delta
\Big)
\stp 0\,.
\eeao
Therefore it suffices to
prove that
\beao
(m/n)^{1/2} \sum_{i=1}^N (S_{K_i,L_i}- p^{-1}(\ov I_n-p_0))\std N(0,\sigma^2(C))
\eeao
conditional on $(X_i)$. Now, an Anscombe type argument
(e.g. Embrechts et al. \cite{embrechts:kluppelberg:mikosch:1997},
Lemma~2.5.8) combined with the \asy\
normality of $((N-np)/\sqrt{np})$ as $np\to\infty$ show that the
random index $N$ in the sum above can be replaced by any integer
\seq\ $\ell=\ell_n\to\infty$ satisfying the relation $np/\ell\to
1$. Given $(X_i)$, the triangular array
\beao
(m/n)^{1/2} (S_{K_i,L_i}- p^{-1}(\ov I_n-p_0))\,,\quad
i=1,\ldots,\ell\,,\quad n=1,2,\ldots\,,
\eeao
consists of row-wise iid mean zero \rv s, hence it satisfies the
assumption of infinite smallness conditional on $(X_i)$. Therefore it suffices to
apply a classical \clt\ for triangular arrays of independent \rv s
conditional on $(X_i)$. In view of \eqref{eq:2a}, $\sigma^2(C)$ is the
\asy\ variance of the converging partial sum \seq . Thus it suffices
to prove the following Lyapunov condition conditional on $(X_i)$:
\beao
(m/n)^{3/2} \ell E^\ast \big(|S_{K_1,L_1}- p^{-1}(\ov I_n-p_0)|^3\big)
&\sim & m^{3/2} n^{-1/2}\,p\, E^\ast \big(|S_{K_1,L_1}- p^{-1}(\ov
I_n-p_0)|^3\big)
\stp 0\,.
\eeao
An application of the $C_r$-inequality yields
\beao
\lefteqn{ m^{3/2} n^{-1/2}\,p\, E^\ast \big(|S_{K_1,L_1}- p^{-1}(\ov
I_n-p_0)|^3\big)}\\
&\le &4 \, m^{3/2}\, n^{-1/2}\,p\,\Big[ E^\ast \big(|S_{K_1,L_1}|^3\big)+p^{-3}|\ov
I_n-p_0|^3
\Big]\\[2mm]
&=& 4 m^{3/2} n^{-1/2}\,p\,E^\ast \big(|S_{K_1,L_1}|^3\big)
+4(n\,p)^{-2} |(n/m)^{1/2}\wt P_m|^3\,.
\eeao
Since the last expression is $o_P(1)$ it suffices to show that
\beam\label{eq:hilfa}
 m^{3/2} n^{-1/2}\,p\,E^\ast \big(|S_{K_1,L_1}|^3\big)\stp 0\,.
\eeam
An application of Markov's inequality shows that it suffices to
switch to unconditional moments in the last expression. Writing
$S_{l}=\wt I_1+\cdots +\wt I_{l}$,
we have
by stationarity of $(X_i)$
\beao
E\big(|S_{K_1,L_1}|^3\big)=E\big(|S_{L_1}|^3\big)
=\sum_{l=1}^\infty E(|S_l|^3) \, (1-p)^{l-1}\,p\,.
\eeao
Next we employ a moment bound due to Rio \cite{rio:1994}, p. 54,
\beao
E(|S_l|^3)\le 3 \wt s_l^{3}+ 144\,l\,\int_0^1
[\alpha^{-1}(x/2)\wedge l]^2\,Q^3(x)\,dx\,,
\eeao
where $\alpha^{-1}$ denotes the generalized inverse of the rate \fct\
$\alpha(t)=\alpha_{[t]}$, $Q$ is the quantile \fct\ of the \ds\ of
$|\wt I_1|=|I_1-p_0|$ and
\beao
\wt s_l^2= \sum_{i=1}^l\sum_{j=1}^l |\cov(I_i,I_j)|
=l \,p_0\,(1-p_0)+2\sum_{h=1}^{l-1}(l-h)\,|p_{0h}-p_0^2|\,.
\eeao
For every fixed $k\ge 1$,
\beao
\wt s_l^2&\le & (l/m) \left[m\,p_0+2 \left(\sum_{h=1}^k+\sum_{h=k+1}^{r_n}+\sum_{h=r_n+1}^{l-1}\right) (1-h/l)\,m\,|p_{0h}-p_0^2|\right]\,,
\eeao
where sums over empty index sets are zero. Using \regvar\ of $(X_i)$
and the mixing condition (M), we conclude that the \rhs\ is of the
order $O(l/m)$ uniformly for $l$. Hence
\beao
m^{3/2}\,n^{-1/2}\,p\,\sum_{l=1}^\infty \wt s_l^3\,(1-p)^{l-1}\,p
&\le &c\,n^{-1/2}\,p\sum_{l=1}^\infty l^{3/2}\,(1-p)^{l-1}\,p\\[2mm]
&\le & c\,(n\,p)^{-1/2}\to 0\,.
\eeao
Direct calculation with the quantile \fct\ of $|\wt I_t|$ shows
that
\beao
\int_0^1 (\alpha^{-1}(x/2)\wedge l)^2 Q^3(x)dx&=&
p_0^3\int_0^{1-p_0} [\alpha^{-1}(x/2)\wedge l]^2dx+(1-p_0)^3
\int_{1-p_0}^1 [\alpha^{-1}(x/2)\wedge l]^2\,dx\\[2mm]
&\le& c\,\big[m^{-3}\sum_{k=1}^\infty k \alpha_k + m^{-1}\big]=O(m^{-1})\,.
\eeao
In the last step we used condition \eqref{eq:10}.
Combining the estimates above, we obtain
\beao
m^{3/2}n^{-1/2}\,p\,\sum_{l=1}^\infty E(|S_l|^3) (1-p)^{l-1}\,p
&\le & c\,[(n\,p)^{-1/2}+ (m/n)^{-1/2}]\to 0\,.
\eeao
This proves relation \eqref{eq:hilfa} and concludes the proof of the theorem.

\subsection{Proof of Theorem~\ref{thm:2}}
From \eqref{e:hilfg} we know that
\beam\label{eq:hilfc}
P^\ast(|\wh P_m^\ast(D_i)- \mu(D_i)|>\delta)\stp 0\,,\quad \delta>0\,,\quad
i=1,\ldots,h+1,
\eeam
therefore \eqref{eq:hilfb} follows.
\par
Relation \eqref{eq:hilfc} implies that
for each $i=1,\ldots,h$, in $P^\ast$-\pro y,
\beao
\wh{\rho}^\ast_{C,D_i}-\wh{\rho}_{C,D_i} &=&
\dfrac{\wh P_m^\ast(D_i)\wh
  P_m(C)-\wh P_m^\ast (C)\wh P_m(D_i)}{\wh P_m^\ast(C)\wh
  P_m^\ast(C)}\\[2mm]
&=& \dfrac{1+o_P(1)}{\mu^2(C)}\Big[ \mu(C) (\wh P_m^\ast(D_i)-\wh
P_m(D_i))
-\mu (D_i) (\wh P_m^\ast(C)-\wh P_m(C) )
\Big] \,.
\eeao
Therefore it suffices for the \clt\ \eqref{eq:cltbs} to prove a multivariate \clt\ for the
quantities $\wh P_m^\ast(D_i)-\wh P_m(D_i)$, $i=1,\ldots,h+1$. We will
show the result for $h=1$; the general case is analogous. It will be
convenient to write $D=D_1$ and $C=D_2$.
\ble
The following central limit theorem holds in $P^\ast$-\pro y
\beao
\bfS_n=(n/m)^{1/2} \left(\barr{ll}
\wh P_m^\ast(D)-\wh P_m(D)\\[2mm]
\wh P_m^\ast(C)-\wh P_m(C)\earr
\right)\std N(\bf0,\Sigma)\,,
\eeao
where the \asy\ covariance matrix is given by
\beao
\Sigma& =&\left(\barr{ll}\sigma^2(D)& r_{DC}\\[2mm]
r_{DC}&\sigma^2(C) \earr\right)\,,\\[2mm]
r_{DC}&=&\mu(C\cap D) +  \sum_{i=1}^\infty{[ \mu_{i+1}(D\times
\ov \bbr_0^{d(i-2)}\times
C)+\mu_{i+1}(C\times
\ov \bbr_0^{d(i-2)}\times
D)]}\,.
\eeao
\ele
\begin{proof}
We show the result by using the Cram\'er-Wold device, i.e.
\beao
\bfz'\bfS_n \std N(0,{\bfz'\Sigma\bfz})\,,\quad \bfz\in \bbr^2\,.
\eeao
We indicate the main steps in the proof in which we
follow the lines of the proof of Theorem~\ref{thm:1}. We observe
that $E^\ast(\bfz'\bfS_n)=0$. Next we show that, conditional on $(X_t)$
\beam\label{eq:variance}
\var^\ast (\bfz'\bfS_n ) &=&(n/m)\Big[z_1^2 \var^\ast(\wh P_m^\ast(D))+
z_2^2 \var^\ast(\wh P_m^\ast(C))+2 z_1z_2 \cov^\ast(\wh
P_m^\ast(C),\wh P_m^\ast(D))\Big]\nonumber\\&\stp& \bfz'\Sigma\bfz\,.
\eeam
By \eqref{eq:2a}, $(n/m)\var^\ast(\wh P_m^\ast(D))\stp
\sigma^2(D)$ and $(n/m)\var^\ast(\wh P_m^\ast(C))\stp
\sigma^2(C)$. Hence it suffices to show that
\beam\label{eq:34}
(n/m)\cov^\ast(\wh
P_m^\ast(C),\wh P_m^\ast(D))\stp r_{DC}\,.
\eeam
We observe that
\beao
\cov^\ast(\wh
P_m^\ast(C),\wh P_m^\ast(D))=\frac 1 4 \Big[
\var^\ast(P_m^\ast(C)+\wh P_m^\ast(D))-
\var^\ast(P_m^\ast(C)-\wh P_m^\ast(D))
\Big]\,.
\eeao
Observe that $P_m^\ast(C)\pm \wh P_m^\ast(D)$ contain the
bootstrap \seq s $I_t^\ast(C)\pm I_t^\ast(D)=(I_t(C)\pm I_t(D))^\ast$, $t=1,\ldots,n$.
Therefore the same ideas as for Lemma~5.2 in Davis and Mikosch
\cite{davis:mikosch:2009} and in the proof of Theorem~\ref{thm:1} above
apply to show \eqref{eq:34}.  We omit the details.
\par
It immediately follows from Lemma~\ref{lem:1} and the argument
following it that the multivariate \clt\ can be reduced to the
\clt\ for the triangular array
\beam\label{eq:hilfd}
&&(m/n)^{1/2} \Big[z_1 \big(S_{K_i,L_i}(D) - p^{-1}(\ov I_n(D)-p_0(D))\big)+
z_2
\big( S_{K_i,L_i}(C) - p^{-1}(\ov I_n(C)-p_0(C))\big)\Big],\nonumber\\&&
i=1,\ldots,\ell\,,\quad n=1,2,\ldots\,,\nonumber\\
\eeam
where $\ell=\ell_n$ satisfies the relation $np/\ell \to 1$.
This array consists of row-wise iid mean zero \rv s,
conditional on $(X_t)$. Relation \eqref{eq:variance} yields the
correct \asy\ variance for the \clt\ of the quantities
\eqref{eq:hilfd}. Therefore it again suffices to apply a Lyapunov
condition of order 3 to the summands \eqref{eq:hilfd} conditional on $(X_t)$.
However, an application of the $C_r$-inequality yields that, up to a
constant
multiple,
this Lyapunov ratio is bounded by the sum of the Lyapunov ratios of
 $S_{K_1,L_1}(C)$ and $S_{K_1,L_1}(D)$ which, conditional on $(X_t)$,
were shown to converge to zero in the proof of Theorem~\ref{thm:1}.
This finishes the sketch of the proof of the theorem.
\end{proof}

\end{document}